\pdfoutput=1

\documentclass[fleqn,usenatbib,useAMS]{mnras}

\usepackage{graphicx}	
\usepackage{amsmath}	
\usepackage{multicol}        
\usepackage{bm}		
\usepackage{pdflscape}	
\usepackage{latexsym}
\usepackage[authoryear]{natbib}
\usepackage{bm,nicefrac}

\usepackage[T1]{fontenc}
\usepackage{ae,aecompl}

\usepackage{newtxtext,newtxmath}
\usepackage{orcidlink}

\usepackage[normalem]{ulem}
\def\deg{\ifmmode^\circ\else$^\circ$\fi}

\def\Q{\ifmmode\mathcal{Q}\else$\mathcal{Q}$\fi}
\def\Mach{\ifmmode\mathcal{M}\else$\mathcal{M}$\fi}

\title[Study gas kinematics around G8.14+0.23]
{Molecular Gas Kinematics and Converging Motions in the Filamentary Cloud Hosting G8.14+0.23}
\author[L.~K. Dewangan et al.]{L.~K. Dewangan$^{1}$\thanks{E-mail: lokeshd@prl.res.in}\orcidlink{0000-0001-6725-0483}, 
O.~R. Jadhav$^{1,2}$\orcidlink{0009-0001-2896-1896},
I.~I. Zinchenko$^{3}$\orcidlink{0000-0003-2793-8229},
D.~K. Ojha$^{4}$\orcidlink{0000-0001-9312-3816},
L.~E. Pirogov$^{3}$\orcidlink{0000-0001-6251-2421}, and
\newauthor
A.~K. Maity$^{5}$\orcidlink{0000-0002-7367-9355}\\
$^{1}$Physical Research Laboratory, Navrangpura, Ahmedabad - 380 009, India.\\
$^{2}$Indian Institute of Technology Gandhinagar Palaj, Gandhinagar 382355, India.\\ 
$^{3}$Institute of Applied Physics of the Russian Academy of Sciences, 46 Ulyanov st., Nizhny Novgorod 603950, Russia.\\
$^{4}$Department of Astronomy and Astrophysics, Tata Institute of Fundamental Research, Homi Bhabha Road, Mumbai 400005, India.\\
$^{5}$Faculty of Engineering, Gifu University, 1-1 Yanagido, Gifu 501-1193, Japan.\\}

\begin{document}

\date{ }

\pagerange{\pageref{firstpage}--\pageref{lastpage}} \pubyear{2023}

\maketitle

\label{firstpage}

\begin{abstract}
To investigate massive star formation (MSF), we present an observational study of an elongated filamentary structure (EFS; extent $>$15 pc) hosting the G8.14+0.23 H\,{\sc ii} region ($d$ $\sim$3.0 kpc), infrared dark features, clusters of young stellar objects (YSOs), and ongoing MSF. New APEX $^{12}$CO(2--1), $^{13}$CO(2--1), and C$^{18}$O(2--1) observations (beam size $\sim$25$''$--28$''$) are combined with archival infrared-to-radio data to investigate the molecular-gas distribution and kinematics. The integrated-intensity maps reveal a filamentary molecular cloud spatially consistent with the EFS, while the H$_2$ column density derived from C$^{18}$O(2--1) yields a total cloud mass of $\sim$10$^{5}$ $M_{\odot}$. Two prominent molecular components are identified toward the EFS in position-position-velocity space. Toward the dark-feature region, they are separated by $\sim$2.4 km s$^{-1}$ and exhibit coherent spatial evolution in the channel maps and position-velocity diagrams, with increasing spatial association toward their overlapping region. These characteristics are consistent with possible low-velocity converging motions and molecular-gas assembly within the larger molecular reservoir. Mid-infrared images (resolution $\sim$6$''$) toward the dark-feature region in the EFS, reveal at least three compact infrared features, surrounded or approached by filamentary structures and associated with YSO clusters, indicating localized star formation within the EFS. In contrast, the G8.14+0.23 H\,{\sc ii} region exhibits a broad velocity spread and a circular velocity structure, consistent with expansion and stellar-feedback influence. Overall, the EFS contains organized, velocity-coherent molecular-gas structures and localized star formation, with possible large-scale gas assembly through low-velocity converging motions and feedback-driven restructuring of the molecular gas around the H\,{\sc ii} region. 
\end{abstract}
%
%
\section{Introduction}
\label{sec:intro}
The formation of massive stars ($>$ 8 $M_{\odot}$) requires the accumulation
of large amounts of molecular gas into dense, gravitationally unstable
structures, yet how this material is assembled and concentrated across
multiple spatial scales prior to the onset of strong stellar feedback
remains poorly understood \citep[e.g.,][]{mckee_2003,tan14,Motte+2018,
semadeni19,padoan20,dewangan24,dewangan26g286,Bhadari_2025}. In particular, the physical processes that transport
and organize molecular material from cloud scales into the dense
environments where massive stars and stellar clusters emerge are not yet
fully established. Characterizing the spatial and velocity structure of
molecular gas during these early stages is therefore essential for understanding the initial conditions of massive star formation (MSF). 

Molecular clouds exhibit complex filamentary and hierarchical structures, and numerical simulations suggest that large-scale converging flows can generate such structures and channel molecular material toward dense regions, potentially linking cloud-scale gas dynamics to the formation of massive stars and stellar clusters \citep{semadeni07,smith16,nozaki26}. These converging motions may arise from turbulent compression, global gravitational collapse, or the self-gravity of filamentary molecular networks \citep{myers09,Motte+2018,padoan20,semadeni19}, and can promote the concentration of molecular gas into progressively denser clumps and cores. Observationally, however, identifying converging flows remains challenging, since projected morphology and velocity gradients can also result from gravitational collapse, rotation, stellar feedback, or the superposition of distinct velocity components \citep[e.g.,][]{dewangan26monr2}. In addition, cloud-cloud collisions can produce multiple velocity components and complementary spatial morphologies, making it important to distinguish converging motions from collision-related kinematics. In this context, the target of this work is the promising massive star-forming region, G8.14+0.23.

Using 21 cm continuum observations from the VLA in the DnC configuration, \citet{kim01} identified G8.14+0.23/IRAS 17599-2148 as a bipolar blister-type H\,{\sc ii} region ($\sim$4\rlap.{$'$}7 $\times$ 3\rlap.{$'$}4) aligned along the north--south direction. This H\,{\sc ii} region is excited by an O-type star \citep[e.g.,][]{kim01,kim03,dewangan16b,dewangan19a}, and is linked to a bipolar nebula (BPN) comprising two prominent mid-infrared (MIR) bubbles, CN107 and CN109 \citep{churchwell07,dewangan12}. The cloud associated with the site G8.14+0.23 was studied in a velocity range of $\sim$12--25 km s$^{-1}$ \citep{takeuchi10,dewangan19a}. On larger scales, infrared and sub-millimetre (mm) data covering a field of $\sim$20\rlap.{$'$}8 $\times$ 13\rlap.{$'$}2 around G8.14+0.23 revealed an elongated filamentary structure (EFS), which appears as an infrared-dark cloud (IRDC) in MIR images \citep[see][for more details]{dewangan16b}. 
Based on the temperature distribution and ionized emission, \citet{dewangan16b} reported that the EFS consists of two environments: cold dust/gas regions with no radio emission and areas related to the BPN associated with the H\,{\sc ii} region. The EFS contains several 870 $\mu$m dust continuum clumps \citep{urquhart18} at a distance of $\sim$3.0 kpc \citep[e.g.,][]{dewangan19a}. One of these clumps, located toward the BPN, is associated with a 6.7 GHz methanol maser emission \citep[MME; peak velocity $\sim$19.83 km s$^{-1}$;][]{szymczak12}, further supporting ongoing MSF. This velocity of MME is consistent with that of the molecular cloud, indicating a physical association. Notably, the 6.7 GHz MME is offset from the radio continuum peaks (i.e., $\sim$0\rlap.{$''$}25) reported in \citet{dewangan16b}. 
Figure~\ref{fg1}a displays the MIR 12 $\mu$m image, showing the previously reported extinction features (i.e., dark regions), the MIR 
bubbles (CN107 and CN109), and the BPN (outlined by the red contour). In Figure~\ref{fg1}a, the star symbol highlights the position of the 6.7 GHz 
MME \citep[from][]{szymczak12}, located toward the waist of the BPN (indicated by the thick cyan line).

Embedded young stellar objects (YSOs) within the EFS, including the BPN, have been identified from near-infrared (NIR) and MIR  observations \citep{dewangan16b,dewangan19a}. Their spatial distribution reveals clustered star formation across the EFS, with some YSO clusters located near the edges of the BPN and potentially influenced by the expanding H\,{\sc ii} region, while others are distributed away from the prominent ionized gas. These results indicate that star formation occurs across the EFS under different environmental conditions. Hence, this target provides a promising laboratory for investigating the initial conditions and physical mechanisms governing the formation of massive stars and stellar clusters. In other words, regions hosting embedded YSOs but lacking prominent ionized gas provide a particularly valuable opportunity to investigate the molecular conditions and gas-assembly processes preceding the development of strong ionizing feedback.
 
Using NANTEN2 $^{13}$CO(1--0) line data (beam size $\sim$200$''$ or $\sim$2.9 pc) covering a 1$\degr$ $\times$ 1$\degr$ area around G8.14+0.23, \citet{dewangan19a} identified two velocity components toward the region and proposed a possible cloud-cloud collision based on their spatial and kinematic association. However, the coarse angular resolution of the NANTEN2 data limits the ability to resolve the detailed molecular-gas structure within the EFS. Higher-resolution molecular-line observations are therefore essential for examining  the detailed gas structure within the EFS. 

Motivated by this, the present work combines new APEX 12-m $^{12}$CO(2--1), $^{13}$CO(2--1), and C$^{18}$O(2--1) observations (beam size $\sim$25$''$--28$''$ or $\sim$0.36--0.41 pc) with archival infrared-to-radio data, including the 1.3 GHz continuum observations of G8.14+0.23. The APEX observations provide substantially finer angular resolution and enable us to characterize the molecular-gas distribution and kinematics across the EFS. In particular, they reveal finer kinematic structure within the broader velocity field identified by \citet{dewangan19a}, allowing us to probe the ongoing physical processes driving the molecular-gas dynamics and large-scale assembly. Specifically, we investigate: (1) how the molecular gas is spatially and kinematically organized across the EFS; (2) whether the identified velocity components exhibit coherent signatures of large-scale converging motions; and (3) how these molecular-gas structures and kinematics are related to the localized star formation and the expanding G8.14+0.23 H\,{\sc ii} region. These analyses allow us to assess the role of large-scale molecular-gas assembly in the ongoing MSF within the EFS.

Section~\ref{sec:obser} outlines the observational data sets used in this study. The results derived from these multi-wavelength data sets are presented in Section~\ref{sec:data}. In Section~\ref{sec:disc}, we discuss the implications of our findings toward the EFS. 
Finally, Section~\ref{sec:conc} provides a summary of our main conclusions.
\section{Data and analysis}
\label{sec:obser}
\subsection{New Observations}
\label{sec:obser1}
Observations with the APEX 12-m radio telescope were performed in July and October 2019 in the framework of the project O-0104.F-9301A-2019 (PI: Lokesh Dewangan). Two frequency setups were used to cover $^{12}$CO, $^{13}$CO and C$^{18}$O $J=2-1$ lines as well as CS and C$^{34}$S $J=5-4$ lines at frequencies from $\sim$220 to $\sim$244 GHz. In both setups, the backend provided a total bandwidth of 16 GHz, with a channel width of 61 kHz, corresponding to a spectral (velocity) resolution of $\sim$0.08 km s$^{-1}$. At these frequencies, the half power beam width ranges from $\sim 25^{\prime\prime}$ to $28^{\prime\prime}$ (or $\sim$0.36 to 0.41 pc), and the main-beam efficiency is $\sim$0.7--0.8. The target region was mapped in the On-The-Fly mode, using a reference position at RA(J2000) = 17$^{h}$58$^{m}$00$^{s}$, DEC(J2000) = $-$20$\degr$44$'$00$''$, which was checked to be free of contaminating emission. The weather conditions were good (PWV in most sessions was from $\sim$1 to $\sim$2 mm). The achieved rms noise per channel is $\sim$0.15--0.2~K on the T$_{mb}$ scale. Data reduction was performed with the GILDAS package\footnote[1]{https://www.iram.fr/IRAMFR/GILDAS}. The lines $^{12}$CO, $^{13}$CO, and C$^{18}$O were observed for a region of $\sim$18\rlap.{$'$}14 $\times$ 13\rlap.{$'$}64, centered at RA(J2000) = 18$^{h}$03$^{m}$09.4$^{s}$, DEC(J2000) = $-$21$\degr$46$'$16\rlap.{$''$}4. Due to time constraints, only a fraction of this area was mapped in the CS and C$^{34}$S $J=5-4$ lines; therefore, these data are not analyzed in the present work. 
\subsection{Other data}
\label{sec:obser2}
This study used a wide range of existing datasets, including 
the 12 $\mu$m image from the Unblurred Coadds of the Wide-field Infrared Survey Explorer Imaging \citep[unWISE; resolution $\sim$6$''$;][]{Lang_2014}, 
the 24 $\mu$m image from the {\it Spitzer} MIPS Inner Galactic Plane Survey \citep[MIPSGAL; resolution $\sim$6$''$;][]{carey05}, 
and the 1.3 GHz radio continuum map from the South African Radio Astronomy Observatory (SARAO) MeerKAT Galactic Plane Survey \citep[SMGPS; resolution $\sim$8$''$;][]{Goedhart_2024}. 

We also employed surface density contours of YSOs identified using infrared photometric data spanning 1.6--24 $\mu$m \citep{dewangan16b}, together with the positions of the APEX Telescope Large Area Survey of the Galaxy \citep[ATLASGAL; beam size $\sim$19\rlap.{$''$}2;][]{schuller09} 870 $\mu$m dust  continuum clumps \citep{urquhart18} and Hi-GAL clumps \citep{elia17}.
\begin{figure}
\center
\includegraphics[width=0.92\linewidth]{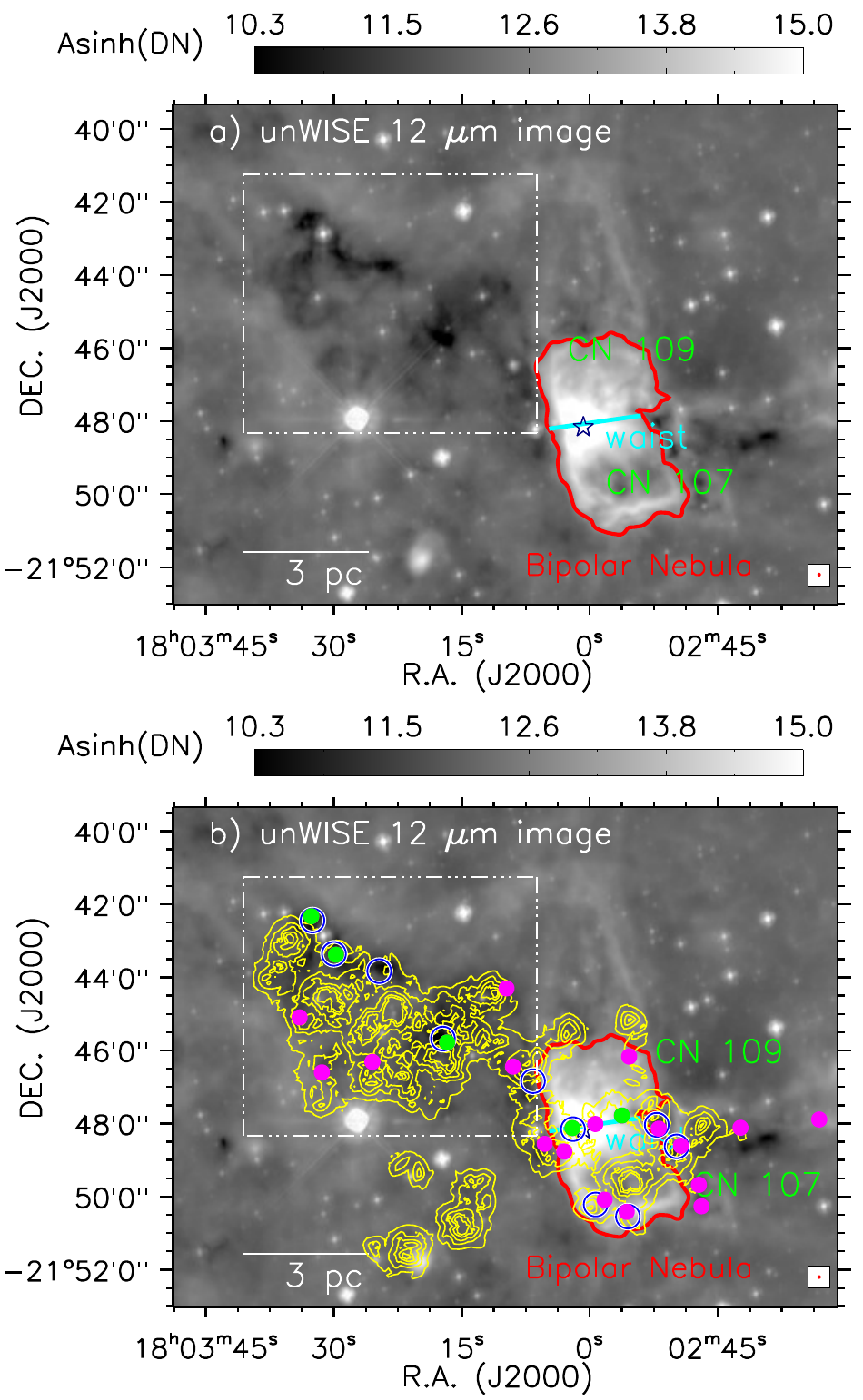}
\includegraphics[width=0.92\linewidth]{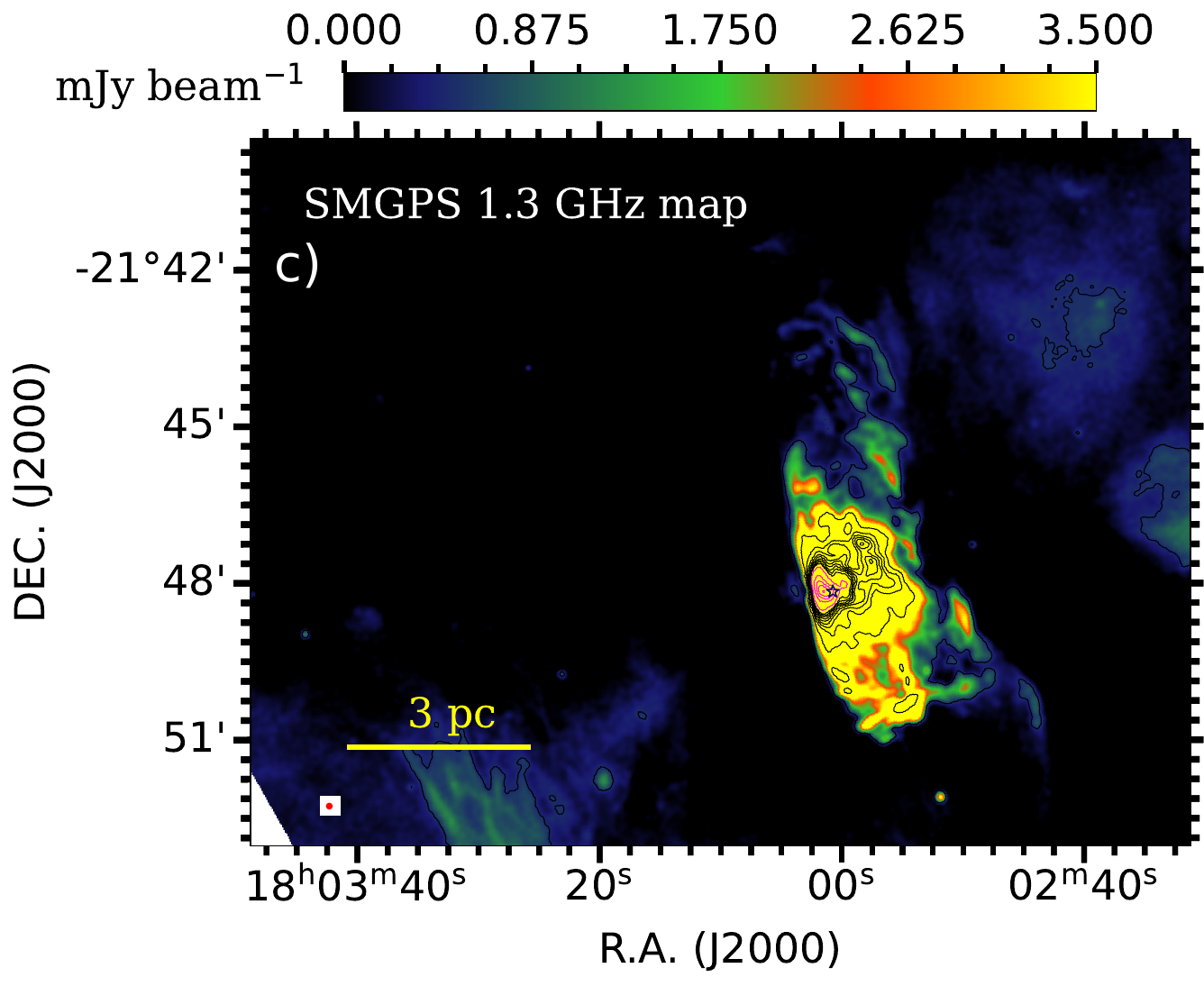}
\caption{a) 12~$\mu$m image (resolution $\sim$6$''$ or $\sim$0.087 pc) of a larger area hosting G8.14+0.23/IRAS 17599-2148 ($\sim$18\rlap.{$'$}14 $\times$ 13\rlap.{$'$}64 or $\sim$15.80 pc $\times$ 11.88 pc at {\it d} $\sim$3.0 kpc), centered at $\alpha_{2000}$ = 18$^{h}$03$^{m}$09\rlap.$^{s}$9, $\delta_{2000}$ = $-$21$\degr$46$'$14\rlap.{$''$}8. The thick red contour (86250 DN) outlines the BPN, with its waist marked by the cyan line. b) Same as panel ``a'', overlaid with YSO surface-density contours of 5, 8, 12, 18, and 25~YSOs~pc$^{-2}$ \citep{dewangan16b}. Open circles mark ATLASGAL 870~$\mu$m clumps. Filled green/magenta circles denote Hi-GAL clumps with surface densities above/below 1~g~cm$^{-2}$, respectively. c) SMGPS 1.3~GHz continuum map (resolution $\sim$8$''$ or $\sim$0.12 pc) with black contours of [0.5, 6, 11.5, 17, 22.5, 28, 33.5, 39, 44.5, 50] and magenta contours of [70, 137.5, 205, 272.5, 340]~mJy~beam$^{-1}$. In panels ``a'' and ``b'', the dot-dashed box highlights the area containing the infrared dark features. In each panel, the star marks the 6.7~GHz MME position. The scale bar corresponds to 3~pc, and the beam is shown by the filled red circle in each panel.}
\label{fg1}
\end{figure}
\begin{figure*}
\center
\includegraphics[width=\textwidth]{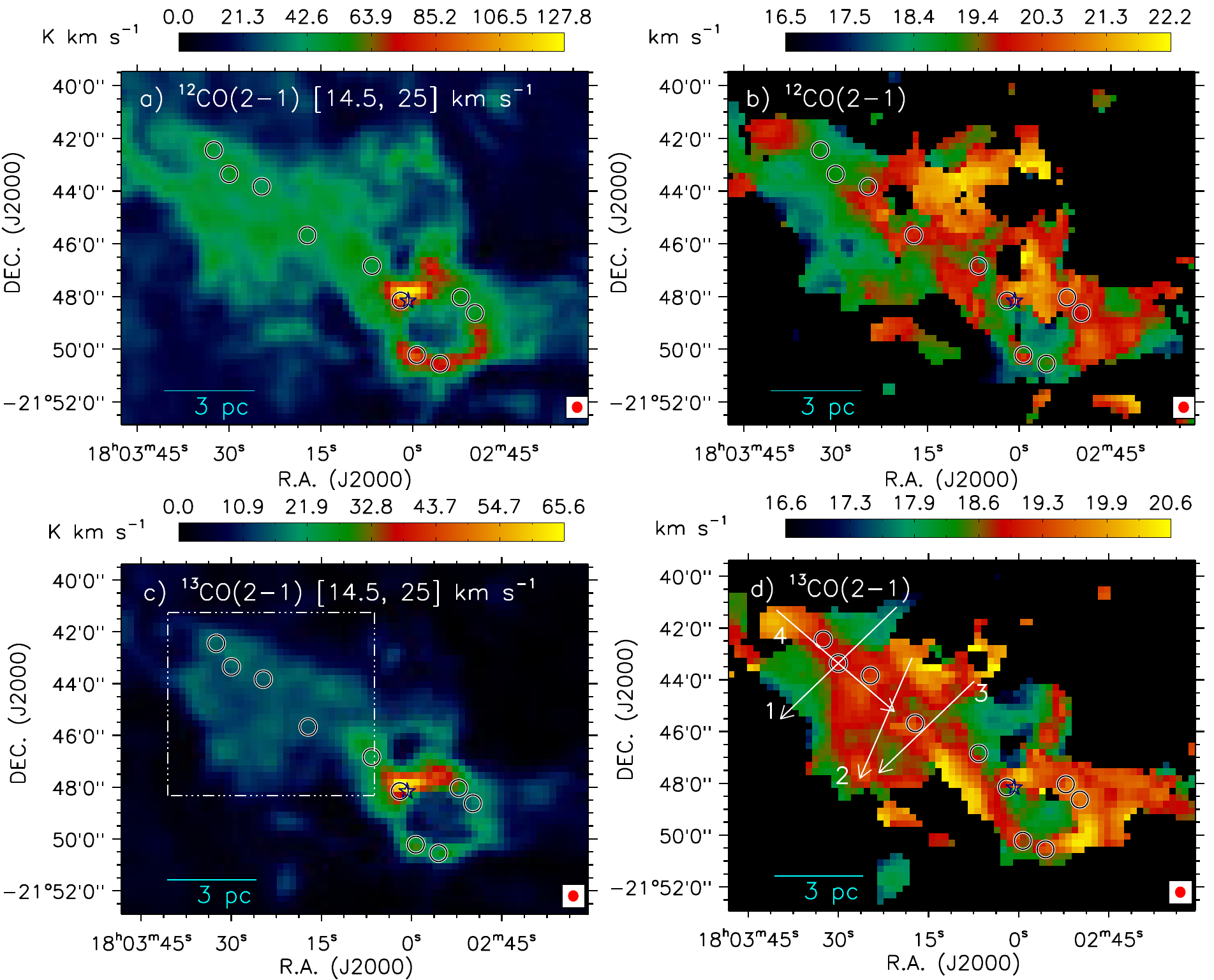}
\caption{a) APEX $^{12}$CO(2--1) moment-0 map integrated over the velocity range of [14.5, 25] km s$^{-1}$. 
b) APEX $^{12}$CO(2--1) moment-1 map.   
c) APEX $^{13}$CO(2--1) moment-0 map at [14.5, 25] km s$^{-1}$. The dot-dashed box is the same as Figure~\ref{fg1}a. d) $^{13}$CO(2--1) moment-1 map. Four arrows (\#1--4) mark the path along which the position-velocity diagrams of the APEX $^{13}$CO(2--1) and C$^{18}$O(2--1) emissions are extracted (see Figures~\ref{fg8} and~\ref{fg9}). In each panel, open circles mark the positions of the ATLASGAL clumps \citep{urquhart18}. The star symbol indicates the position of the 6.7 GHz MME \citep{szymczak12}, and the scale bar corresponds to 3 pc at a distance of 3.0 kpc. The beam size of the APEX line data is shown by a filled red circle in the bottom-right corner.}
\label{fg2}
\end{figure*}
\begin{figure*}
\center
\includegraphics[width=\textwidth]{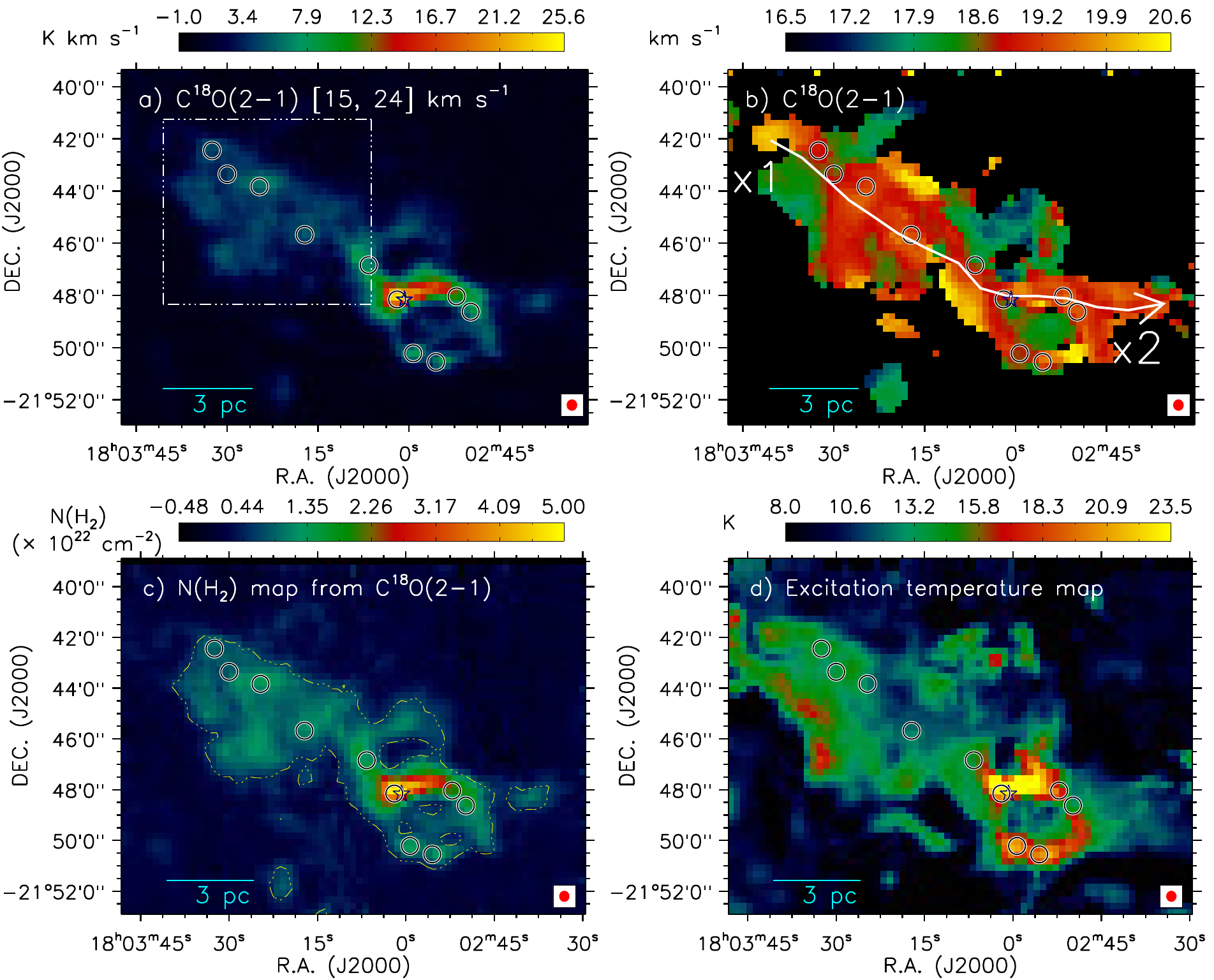}
\caption{a) APEX C$^{18}$O(2--1) moment-0 map integrated over the velocity range of [15, 24] km s$^{-1}$. The dot-dashed box is the same as Figure~\ref{fg1}a. b) APEX C$^{18}$O(2--1) moment-1 map. 
The curve x1--x2  marks the path along which the position-velocity diagrams of the APEX $^{12}$CO(2--1), $^{13}$CO(2--1), and C$^{18}$O(2--1) emissions are extracted (see Figure~\ref{fg4}). 
c)  $N({{\rm{H}}}_{2})$ map derived from the APEX C$^{18}$O(2--1) emission. 
The $N({{\rm{H}}}_{2})$ contour at 5.7 $\times$ 10$^{21}$ cm$^{-2}$ (or 5$\sigma$) is overlaid on the map. d) Excitation temperature map derived from the APEX $^{12}$CO(2--1) emission. 
In each panel, open circles represent the positions of the ATLASGAL clumps \citep{urquhart18}. The star symbol shows the position of the 6.7 GHz MME \citep[from][]{szymczak12}, 
and the scale bar corresponds to 3 pc at a distance of 3.0 kpc. The beam size of the APEX line data is 
indicated by a filled red circle in the bottom-right corner.}
\label{fg3}
\end{figure*}
\begin{figure*}
\center
\includegraphics[width=11cm]{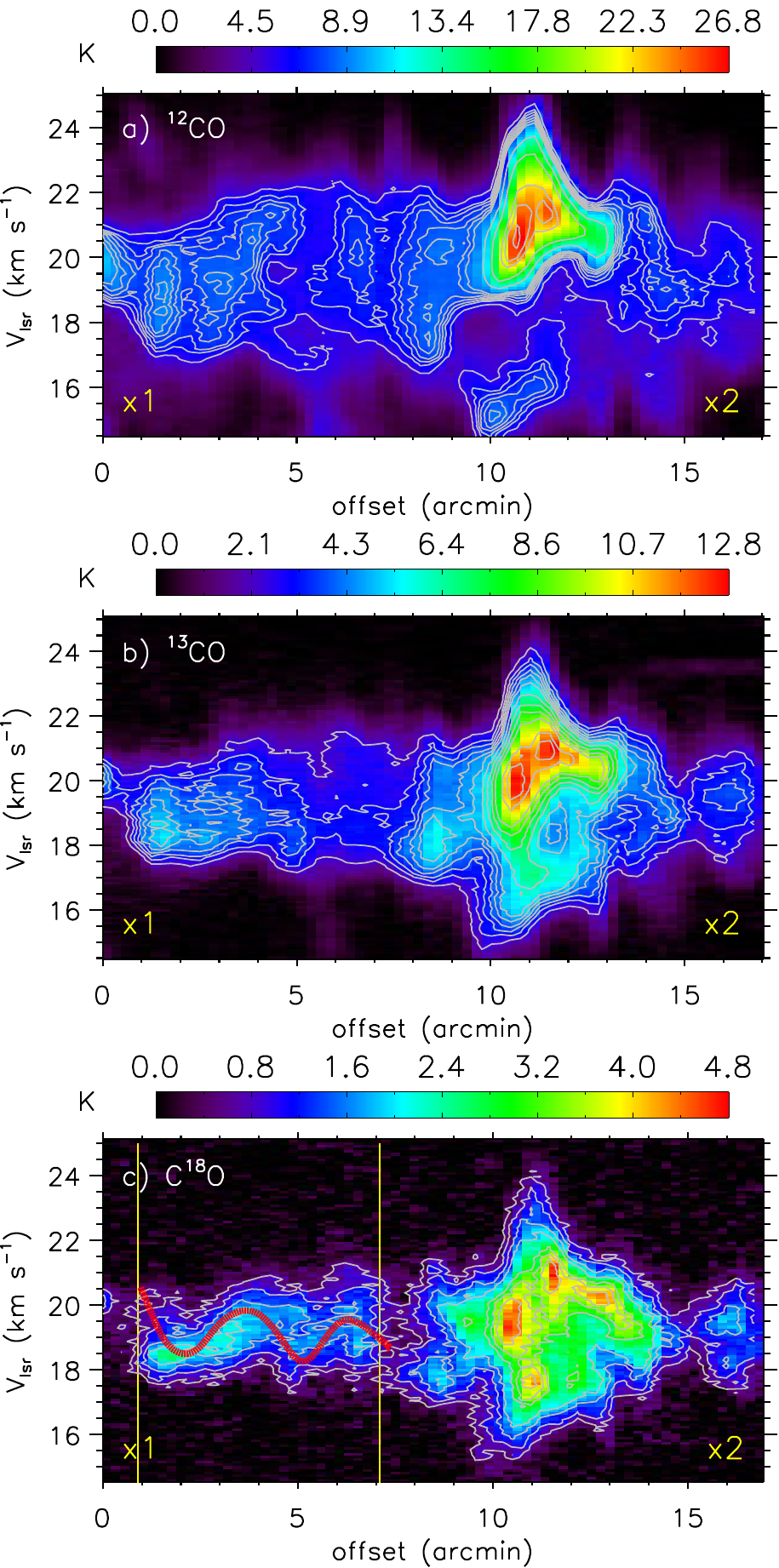}
\caption{a) Position-velocity diagram of the APEX $^{12}$CO(2--1) emission extracted along the curve x1--x2 indicated in Figure~\ref{fg3}b. The $^{12}$CO(2--1) contours (in gray) are also overlaid at levels of 5.5, 6.5, 7, 7.5, 8, 8.5, 9, 12, 16, 22, 24, and 26 K.
b) Same as Figure~\ref{fg4}a, but for the APEX $^{13}$CO(2--1) emission. The corresponding contours (in gray) are plotted at 2.3, 3, 3.5, 4, 4.5, 5, 6, 7, 8, 9, 11, and 12 K. 
c) Same as Figure~\ref{fg4}a, but for the APEX C$^{18}$O(2--1) emission. 
The C$^{18}$O(2--1) contours (in gray) are also shown at levels of 0.6, 1, 1.5, 2, 3, 3.5, and 4 K. The thick red curve highlights the undulating distribution of C$^{18}$O emission in position-velocity space over offsets of approximately 1--7 arcmin (see the vertical lines), suggestive of a velocity oscillation along the structure. 
Each diagram is extracted along a one-pixel-wide path.} 
\label{fg4}
\end{figure*}
\begin{figure*}
\center
\includegraphics[width=\textwidth]{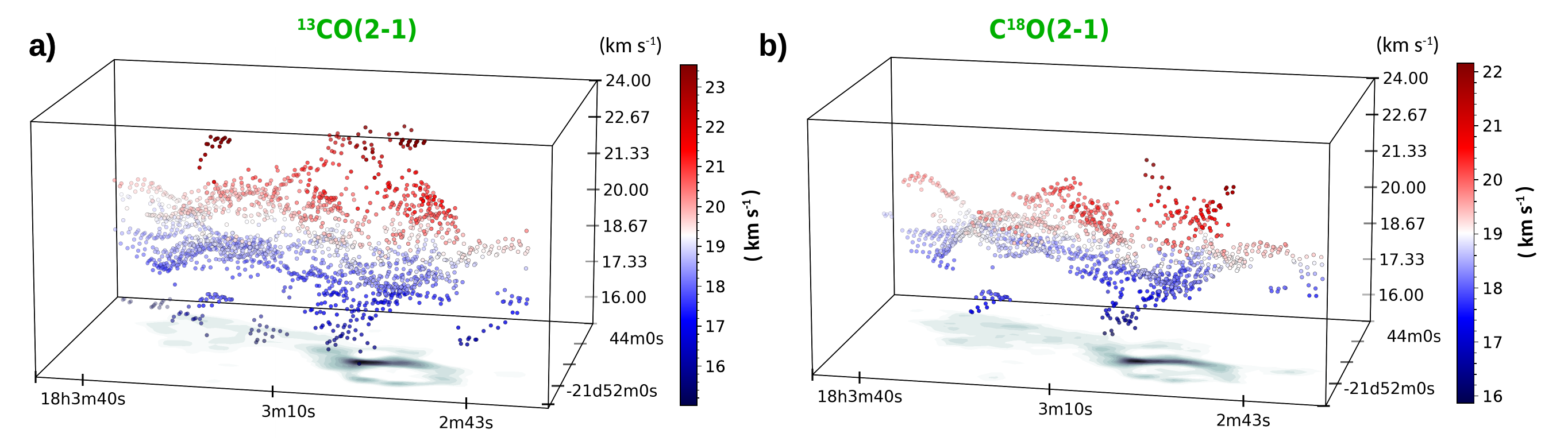}
\caption{a) PPV diagram of the APEX $^{13}$CO(2--1) emission. 
b) Same as Figure~\ref{fg5x}a, but for the APEX C$^{18}$O(2--1) emission.
In each panel, the moment-0 map is shown in the RA-DEC plane.} 
\label{fg5x}
\end{figure*}
\section{Results}
\label{sec:data}
\subsection{Physical structure of the EFS}
\label{ssec:data21}
In the direction of EFS (extent $>$ 15 pc), the unWISE 12 $\mu$m image reveals prominent 
dark features with noticeable substructures along with the previously reported BPN (see Figure~\ref{fg1}a). 
Generally, the dark features obscure background infrared emission, indicating the presence of dense and cold gas and dust \citep[e.g.,][]{peretto16}. 
In Figure~\ref{fg1}a, the red contour overlaid on the unWISE 12 $\mu$m image outlines the BPN, while its waist is 
indicated by the thick cyan line \citep[see][for more details]{dewangan12}. The BPN's lobes extend outward in opposite directions from the waist. The unWISE reprocessing of {\it Wide-field Infrared Survey Explorer} (WISE) data maintains the native $\sim$6$''$ (or $\sim$0.087) resolution, thereby allowing a clearer view of extended features that were otherwise smoothed in the original pipeline, and thus improving the visibility of these structures.

As mentioned, \citet{dewangan16b} investigated the spatial distribution of embedded YSOs toward the EFS and constructed a YSO surface density map (see their paper for details), which enabled the identification of clustered YSO populations. The resulting YSO surface density contours, derived assuming a distance of 3.0 kpc, are overlaid on the unWISE 12 $\mu$m image (see Figure~\ref{fg1}b). In Figure~\ref{fg1}b, the surface density contours (5--25 YSO pc$^{-2}$) from the previous 
study are shown solely to highlight the locations of star formation activity within the EFS.
The positions of 10 ATLASGAL 870 $\mu$m dust continuum clumps located at d $\sim$3.0 kpc \citep{urquhart18} and the 6.7 GHz MME are also highlighted in Figure~\ref{fg1}b. 
Additionally, the positions of 22 Hi-GAL clumps \citep[from][]{elia17} are marked in Figure~\ref{fg1}b.
The Hi-GAL clumps \citep{elia17} were identified from multi-wavelength {\it Herschel} data, with dust temperature, column density, mass, luminosity, and surface density derived from $\sim$36$''$ resolution maps using modified blackbody fitting. 
In our selected target area, the Hi-GAL clumps span masses of $\sim$7.3--114.1~$M_{\odot}$ (median $\sim$24.5~$M_{\odot}$), bolometric luminosities of $\sim$3.3--6015.7~$L_{\odot}$ (median $\sim$527.2~$L_{\odot}$), and surface densities of $\sim$0.12--2.12~g~cm$^{-2}$. 
At least five clumps have surface densities $>1$ g cm$^{-2}$ (see filled green circles in Figure~\ref{fg1}b), reaching the commonly adopted high-surface-density regime associated with MSF \citep[see also][for more details]{elia17}. Three of these clumps lie within the dot-dashed box enclosing the infrared-dark features, indicating dense environments conducive to gravitational collapse and making them potential sites of MSF.

Figure~\ref{fg1}c displays the SMGPS 1.3 GHz continuum emission map (resolution $\sim$8$''$ or $\sim$0.12 pc; rms $\sim$10--20 $\mu$Jy beam$^{-1}$) overlaid with the SMGPS 1.3 GHz contours (in black and magenta). In the direction of the BPN, the extended radio continuum emission appears more prominent in the higher-resolution SMGPS 1.3 GHz map, which may seem unexpected, as higher resolution generally highlights small-scale structures. However, the combination of high sensitivity and improved uv-coverage of SMGPS enables the recovery of both compact and moderately extended emission. As a result, diffuse structures are detected with enhanced clarity and contrast, whereas in lower-resolution data such emission may be diluted or confused with the background noise. The peak of the radio continuum emission, likely tracing the ionizing source(s), is located near one edge of the BPN waist rather than at its geometric center. This configuration differs from the more centrally located ionizing sources reported for the bipolar H\,{\sc ii} region W40 \citep{kuhn10,mallick13}.

Using the {\it clumpfind} IDL algorithm \citep{williams94}, we determined the integrated flux density of 6.37 Jy for the ionized structure traced at 0.5 mJy beam$^{-1}$ (i.e., BPN) from the SMGPS 1.3 GHz continuum map. This corresponds to a Lyman continuum photon rate (log $N_{\rm UV}$ in s$^{-1}$) of 48.65. Based on this estimate, the ionized structure toward BPN is likely powered by an O8V-O7.5V type star \citep{panagia73}. The calculation is carried out assuming an electron temperature of 10$^{4}$~K and a distance of 3 kpc. Further details of the methodology are provided in \citet{dewangan17a} (see also \citealt{matsakis76} for the relevant equation).
\subsection{Molecular line emission toward EFS}
\label{sec:gasmorphb}
In order to study the physical and kinematic properties of the cloud and its associated structures, we produced moment-0 map, moment-1 map, position-velocity diagrams, and position-position-velocity (PPV) diagrams using molecular line data (i.e, $^{12}$CO(2--1), $^{13}$CO(2--1), and C$^{18}$O(2--1); spectral resolution $\sim$0.08 km s$^{-1}$; spatial resolution $\sim$25$''$--28$''$ or $\sim$0.36--0.41 pc) obtained with the APEX 12-m radio telescope. 
\subsubsection{Distribution of molecular gas toward EFS}
\label{zzsec:gas}
The moment-0 map, also referred to as the integrated intensity map, represents the total line emission integrated over a specified velocity range, highlighting the spatial distribution of the molecular gas. Figures~\ref{fg2}a and~\ref{fg2}c present the moment-0 maps of APEX $^{12}$CO(2--1) and APEX $^{13}$CO(2--1), respectively, both integrated in the velocity range of [14.5, 25] km s$^{-1}$. In Figure~\ref{fg3}a, we display the moment-0 map of APEX C$^{18}$O(2--1), which is constructed over a slightly narrower velocity range of [15, 24] km s$^{-1}$. The velocity ranges for each transition are determined from the inspection of their average spectra (not shown). The narrower range adopted for C$^{18}$O ensures robust detection of dense gas emission while minimizing noise contamination, whereas the broader ranges for $^{12}$CO and $^{13}$CO encompass their more extended velocity structures. 
Since C$^{18}$O emission is more optically thin compared to $^{12}$CO and $^{13}$CO, it provides a better probe of the denser molecular gas. In all moment-0 maps, infrared-dark features within the EFS show the presence of dense gas. The BPN stands out as a particularly prominent feature, with localized concentrations of molecular material distributed along its edges and waist. 

The moment-1 map, or intensity-weighted velocity map, provides the mean velocity of the emitting gas at each position on the plane of the sky. Figures~\ref{fg2}b,~\ref{fg2}d, and~\ref{fg3}b display moment-1 maps of APEX $^{12}$CO(2--1), $^{13}$CO(2--1), and C$^{18}$O(2--1), respectively. Across all the moment-1 maps, systematic velocity variations are evident along the EFS and across the BPN, pointing to complex gas motions. In the moment-1 maps, two prominent substructures are evident, distinguished by red and green colors, with a small velocity separation between them.  

Using the distribution of the dense gas tracer C$^{18}$O, we derived the molecular hydrogen column density [$N({{\rm{H}}}_{2})$] map from the emission of APEX C$^{18}$O(2--1). Figure~\ref{fg3}c shows the resulting $N({{\rm{H}}}_{2})$ map, with an overlaid contour corresponding to $N({{\rm{H}}}_{2})$ = 5.7 $\times$ 10$^{21}$ cm$^{-2}$ (or 5$\sigma$). The total column density enclosed within this contour is used to estimate the cloud mass (see Section~\ref{1zzsec:gas}). The detailed methodology adopted to convert the C$^{18}$O emission into column density, including assumptions on excitation conditions and abundance ratios, is described in Section~\ref{1zzsec:gas}. 
In this connection, the excitation temperature ($T_\mathrm{ex}$) map of the target region is derived using APEX $^{12}$CO(2--1) observations (see Section~\ref{1zzsec:gas}), and is presented in Figure~\ref{fg3}d. The $^{12}$CO(2--1) line is used for this purpose as it is generally expected to be optically thick in molecular clouds, making it a reliable tracer for estimating the excitation temperature. Along the EFS, variations in excitation temperature are apparent. 
\subsubsection{Mass of the molecular cloud}
\label{1zzsec:gas}
We estimated the mass of the cloud associated with the EFS using APEX C$^{18}$O(2--1) molecular line emission. This line is generally considered to be optically thin due to the low abundance of C$^{18}$O. However, this assumption may break down in the densest regions, where optical depth effects can lead to an underestimation of column densities. Therefore, while C$^{18}$O(2--1) is commonly used as a tracer of molecular gas, the derived quantities in this work should be interpreted with this caveat in mind. Assuming that the system is in local thermodynamic equilibrium (LTE), the column density of a linear molecule is given by the equation \citep{Magnum_2015}
\begin{eqnarray}
\label{eq1}
N(\rm C^{18}O)&=&\frac{3h}{8\pi^3\mu^2S}\frac{Q_\mathrm{rot}}{g_{J}}
\frac{\mathrm{exp}\left(\frac{E_\mathrm{up}}{k T_\mathrm{ex}}\right)}{\mathrm{exp}\left(\frac{h\nu}{k T_\mathrm{ex}}\right) - 1} \nonumber \\
& &\times\frac{1}{J(T_\mathrm{ex})-J(T_\mathrm{bg})}\int T_\mathrm{mb}~dv,
\end{eqnarray}
where $J(T) = \frac{h\nu/k}{\mathrm{exp}(h\nu/k T)-1}$, $T_{\rm bg}$ = 2.73~K is the cosmic microwave background temperature and $T_\mathrm{ex}$ is the excitation temperature of the target region. The symbol $\mu$ stands for the dipole moment of the molecule. The degeneracy ($g_\mathrm{J}$) is $2J_u+1 = 5$, and the line strength ($S$) is $\frac{J_u}{2J_u+1} = \frac{2}{5}$, where $J_u$ is the rotational quantum number of the upper state (i.e., $J_u = 2$ for the $J = 2-1$ transition). $E_\mathrm{up}$ is the energy of the upper state. We have adopted an approximate formula for the rotational partition function ($Q_\mathrm{rot}$) = $\frac{kT}{hB}~+~\frac{1}{3}$ \citep{Magnum_2015}, where $B$ is the rotational constant of the molecule. From the NASA Jet Propulsion Laboratory database\footnote[2]{https://spec.jpl.nasa.gov}\citep{Pickett_1998} we obtained B = 54891.420 MHz, dipole moment $\mu$ = 0.11079 D and $E_{\rm up}$ = 15.8 K. Assuming LTE and a beam filling factor close to unity, the $T_\mathrm{ex}$ map (see Figure~\ref{fg3}d) is estimated using the following equation \citep{Garden_1991} 
\begin{equation}
T_{\mathrm{ex}} = \frac{11.06}{\ln \left( 1 + \frac{11.06}{T_{\mathrm{peak}} + 0.19} \right)},
\end{equation}
where $T_{\rm peak}$ represents the maximum $T_{\rm mb}$ value of the $^{12}$CO(2--1) emission along the velocity axis at each pixel. We integrated the C$^{18}$O(2--1) emission ($T_{\rm mb}$) in the velocity range [15, 24] km s$^{-1}$ and substituted in equation~\ref{eq1}, to obtain the $N(\rm C^{18}O)$ map. In order to obtain the $N(\rm H_2)$ map, we used the column density ratio between H$_2$ and C$^{18}$O (i.e., $\frac{N(\mathrm{H_2})}{N(\mathrm{C^{18}O})}$ = $3.54\times 10^6$) at a galactocentric distance of 5.2 kpc \citep{Liu_2013}. 
This value accounts for the expected variation in isotopic abundance due to Galactic gradients. However, we note that this ratio may be affected by local conditions such as chemical fractionation, selective photodissociation, and depletion of dust grains, which can introduce uncertainties in the derived H$_2$ column densities. Therefore, the estimated values should be considered subject to these systematic uncertainties. Using the $N(\rm H_2)$ map (see Figure~\ref{fg3}c), we can compute the mass of the cloud using the equation \citep[e.g.,][]{dewangan16b}:
 \begin{equation}
\label{eq3}
M = \mu_{H_2} m_{\rm H} a_{\rm pixel} \Sigma N(\rm {H_2})
\end{equation}
where $\mu_{H_2}$ is the mean molecular weight per hydrogen molecule, taken as 2.8, $m_{\rm H}$ is the mass of the hydrogen atom, and $a_{\rm pixel}$ is the area subtended by each pixel in cm$^2$. We only included pixels that had $N(\rm {H_2})$ values above 5.7 $\times$ 10$^{21}$ cm$^{-2}$ (or 5$\sigma$), and substituted in equation~\ref{eq3}. The total mass of the molecular cloud associated with EFS is $\sim$10$^5$ $M_{\odot}$. The uncertainty in the calculated mass primarily arises from the uncertainty in the derived $N(\rm {H_2})$ values, which can be as high as $\sim$50$\%$  \citep[e.g.,][]{bhadari22}. 
\subsubsection{Position-velocity diagrams}
\label{n3zzsec:gas}
Position-velocity diagrams are constructed to examine the velocity structure toward the cloud associated with the EFS. These diagrams are derived for the APEX $^{12}$CO(2--1), $^{13}$CO(2--1), and C$^{18}$O(2--1) emission along the curve x1--x2, as marked in Figure~\ref{fg3}b. This curve almost outlines the spine of the EFS, including the waist of the BPN (see Figure~\ref{fg1}a). Figures~\ref{fg4}a and~\ref{fg4}b show the position-velocity diagrams of APEX $^{12}$CO(2--1) and $^{13}$CO(2--1), respectively, while Figure~\ref{fg4}c presents the corresponding diagram for APEX C$^{18}$O(2--1). 
The emission contours are overlaid in each panel of Figure~\ref{fg4}. All three tracers show a significantly larger velocity spread toward the BPN's waist (i.e., 7--9 km s$^{-1}$) compared to other parts of the EFS's spine (i.e., 3--5 km s$^{-1}$). The position-velocity diagrams of the $^{13}$CO(2--1) and C$^{18}$O(2--1) emission along the spine of the EFS, excluding the BPN, mainly cover the velocity range V$_{\rm lsr}$ $\sim$17.5--21 km s$^{-1}$. These diagrams suggest that the EFS spine is largely characterized by a coherent velocity structure, as indicated by the relatively narrow velocity range ($\sim$17.5--21 km s$^{-1}$) traced by $^{13}$CO and C$^{18}$O. 

%
%
There is a hint of two closely spaced velocity features ($\sim$2--3 km s$^{-1}$; see Figure~\ref{af1} and Appendix~\ref{sec:posbfield}); however, these features are not clearly distinguishable along all parts of the EFS spine. In the left portions of the position-velocity diagrams, gradual transitions between the two velocities are apparent, suggesting an undulating velocity pattern (see Figures~\ref{fg4}b and~\ref{fg4}c). In particular, the C$^{18}$O position-velocity diagram in Figure~\ref{fg4}c reveals a noticeable undulating distribution of emission over offsets of approximately 1--7 arcmin, highlighted by the thick red curve. The emission exhibits alternating excursions towards higher and lower \(V_{\rm lsr}\), indicating systematic velocity variations along the molecular structure. This morphology further supports the presence of a velocity oscillation along the molecular structure, rather than a uniform velocity field.  
In contrast, the emission at offsets of approximately 9--14 arcmin shows a broader velocity 
distribution/spread (i.e., $\sim$7--9 km s$^{-1}$) toward the waist of the BPN, with multiple intensity peaks spanning a relatively wide velocity range. A prominent circular or arc-shaped velocity structure is also evident toward the BPN. Using JCMT $^{12}$CO(3--2) data, \citet{dewangan16b} reported a C-like or ring-like structure in the position-velocity diagrams of G8.14+0.23/BPN. They interpreted this as evidence of an expanding H\,{\sc ii} region in G8.14+0.23/BPN, with a gas expansion velocity of $\sim$4.5 km s$^{-1}$. In general, position-velocity diagrams of expanding bubbles exhibit circular or elliptical morphologies \citep{Butterfield_2018}, which are well-known signatures of expanding molecular shells \citep[see also][]{arce11}. Consistent with this scenario, the position-velocity diagrams of APEX molecular data toward the G8.14+0.23 H\,{\sc ii} region reveal a clear circular velocity structure surrounded by relatively lower-intensity molecular emission. 
\subsubsection{PPV diagrams of APEX C$^{18}$O(2--1) and $^{13}$CO(2--1) emission}
\label{4zzsec:gas}
We analyzed the PPV distribution of the molecular gas to study its three dimensional (in the plane of the sky and radial velocity) distribution. To quantify the velocity distribution in detail, we performed a spectral decomposition of the data cubes of APEX C$^{18}$O(2--1) and $^{13}$CO(2--1) using the Python-based tool \texttt{SCOUSEPY}\footnote[3]{https://scousepy.readthedocs.io/en/latest} \citep{Henshaw_2016,Henshaw_2019}. This approach allows for a systematic, pixel-by-pixel fitting of spectra across the cloud, enabling us to identify and separate multiple velocity components along each line of sight. 

The decomposition procedure starts by defining the kernel size--referred to as the Spectral Averaging Area (SAA)--which was taken as 3 $\times$ 3 pixels. 
The SAAs are spatially distributed to cover the entire molecular emission region. From each SAA, we extract an averaged spectrum and fit it with multiple Gaussian components, where the number of components is guided by the identified velocity features. These serve as initial guesses for fitting the individual spectra of every pixel within the SAA, ensuring consistent identification of velocity components across the field. Details of the decomposition procedure, representative spectral fits, and distributions of the fitted parameters are provided in Appendix~\ref{ccsec:posbfield}. 

Once the decomposition is completed, we visualize the results by mapping the centroid velocities in PPV space. 
Figures~\ref{fg5x}a and~\ref{fg5x}b show the PPV diagrams of the APEX $^{13}$CO(2--1) and C$^{18}$O(2--1) emissions, respectively, covering the entire velocity range of [15, 24] km s$^{-1}$. The decomposition identifies two distinct velocity components in both $^{13}$CO and C$^{18}$O emission toward the EFS. The spatial distributions and overlap of two velocity components are presented in Figures~\ref{af1b} and~\ref{af1c} (see also Section~\ref{sec:moldt}). As noted earlier, moment-1 maps hint at the presence of two prominent substructures with a small velocity separation toward the EFS (see Figures~\ref{fg2}b,~\ref{fg2}d, and~\ref{fg3}b), and the PPV diagrams further support this interpretation. 
\subsection{Molecular Gas Kinematics toward the Dark Features in the EFS}
\label{sec:moldt}
\begin{figure*}
\center
\includegraphics[width=\textwidth]{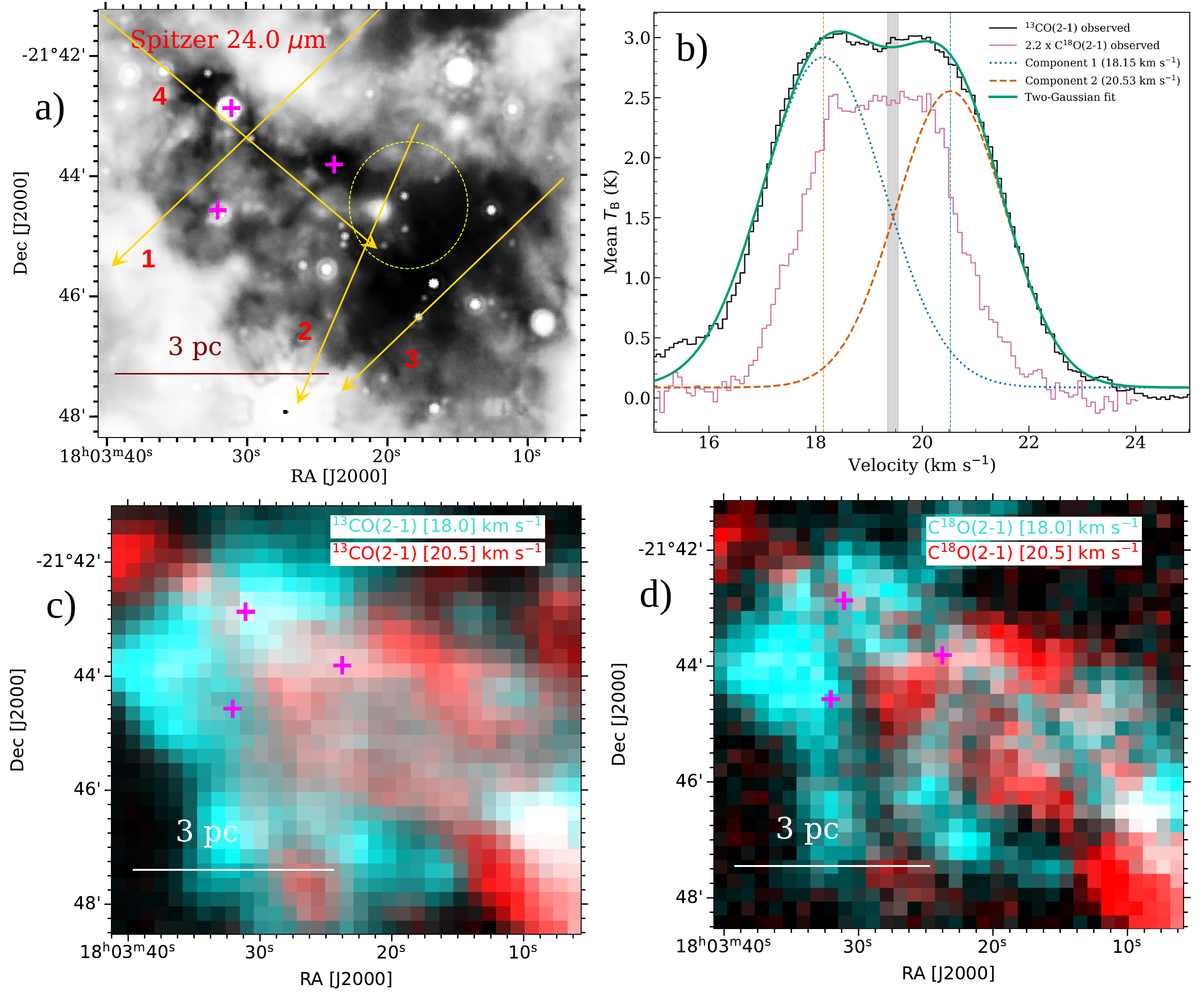}
\caption{a) Zoomed-in {\it Spitzer} 24~$\mu$m view of the region highlighted by the dot-dashed box in Figure~\ref{fg1}a. The dashed circle marks the area from which the APEX $^{13}$CO(2--1) and C$^{18}$O(2--1) spectra are extracted (see Figure~\ref{fg5}b). Four arrows (\#1--4) are the same as those shown in Figure~\ref{fg2}d and indicate the paths along which the position-velocity diagrams are extracted.  
b) The APEX $^{13}$CO(2--1) spectrum toward the region marked by the dashed circle in panel ``a'' is shown along with the Gaussian-fitted profiles. The APEX C$^{18}$O(2--1) spectrum toward the same region is also presented. The two fitted Gaussian components intersect at approximately 19.45 km s$^{-1}$, and the velocity interval of 19.35--19.55 km s$^{-1}$, encompassing the intersection velocity, is highlighted by the gray shaded area. c) Two-color composite map of the APEX $^{13}$CO(2--1) emission, with red and turquoise representing the 20.5 and 18.0~km~s$^{-1}$ velocity channels, respectively. d) Same as panel ``c'', but for the APEX C$^{18}$O(2--1) emission. In panels ``a'', ``c'', and ``d'', plus symbols mark the centers of at least three infrared-dark features. The scale bar corresponds to 3~pc at a distance of 3.0~kpc.}
\label{fg5}
\end{figure*}

In this section, the dark features identified in the 12~$\mu$m image (resolution $\sim$6$''$ or $\sim$0.087 pc) are further examined using the APEX molecular-line data to investigate the kinematic properties of the surrounding gas (see the dot-dashed box in Figures~\ref{fg1}a and~\ref{fg2}c). These dark features are also evident in the {\it Spitzer} 24~$\mu$m image (see Figure~\ref{fg5}a). The MIR images reveal several small-scale extinction filaments (lengths $<2$~pc) that appear to converge morphologically toward compact central regions (see the features surrounding the plus symbols in Figure~\ref{fg5}a), where no radio continuum emission is detected. Since the available molecular-line data have a lower angular resolution of $\sim$25$''$--28$''$, the analysis probes the gas kinematics on spatial scales larger than the individual dark features and does not resolve their internal structure.

Figure~\ref{fg5}b shows the APEX $^{13}$CO(2--1) and C$^{18}$O(2--1) spectra extracted from the region enclosed by the dashed circle in Figure~\ref{fg5}a, along with their respective two-Gaussian fits. This aperture was selected based on the APEX moment-1 maps, which indicate the presence of two distinct molecular structures. Furthermore, the selected region is spatially offset from the G8.14+0.23 H\,{\sc ii} region, thereby minimizing potential contamination from its expanding ionized gas. The two velocity components identified in the aperture-averaged spectra are consistent with those recovered from the spectral decomposition across the entire molecular cloud complex (see Figures~\ref{af1b} and~\ref{af1c} in Appendix~\ref{ccsec:posbfield}), supporting their identification as large-scale velocity structures.

In Figure~\ref{fg5}b, the $^{13}$CO(2--1) spectrum exhibits two prominent velocity components centered at $V_{\rm lsr}$ $\sim$18.1 and 20.5~km~s$^{-1}$, as indicated by the Gaussian fits. The two fitted Gaussian profiles intersect at approximately 19.45 km s$^{-1}$, which provides a basis for defining the velocity ranges of the two cloud components (see Appendix~\ref{ccsec:posbfield} for details). The C$^{18}$O(2--1) spectrum also shows emission over these velocity ranges. Both molecular tracers thus reveal two velocity components separated by approximately 2.4 km s$^{-1}$. This relatively small velocity separation suggests that the two components are closely spaced in velocity and may represent physically related structures within the molecular environment of the EFS. This interpretation is further supported by the PPV diagrams of the APEX $^{13}$CO(2--1) and C$^{18}$O(2--1) emission (Figures~\ref{fg5x}a and~\ref{fg5x}b).

\begin{figure*}
\center
\includegraphics[width=\textwidth]{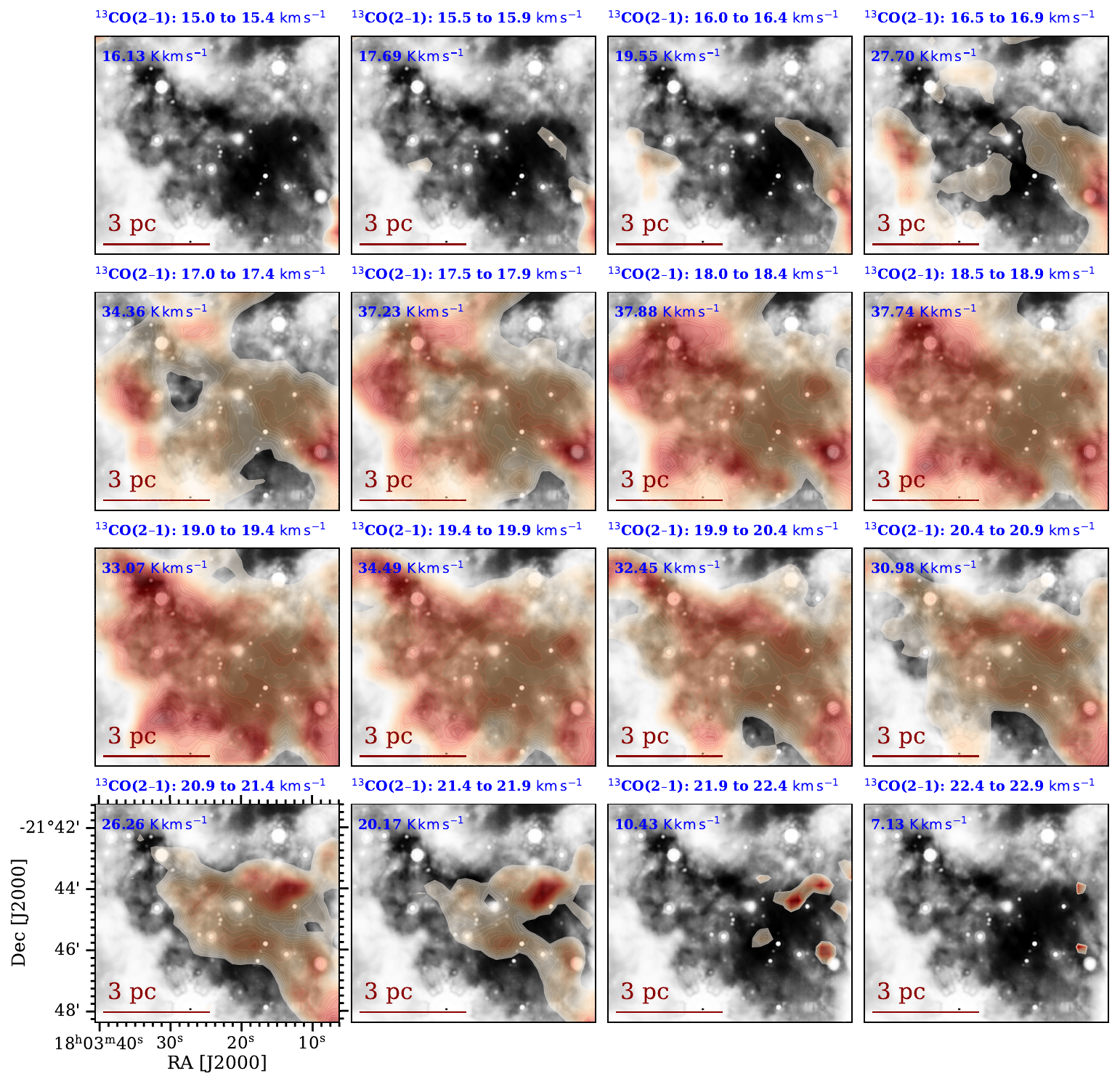}
\caption{Channel maps of the APEX $^{13}$CO(2--1) emission overlaid on the zoomed-in {\it Spitzer} 24~$\mu$m image (see Figure~\ref{fg5}a). 
The $^{13}$CO(2--1) emission is shown as filled contours, with 20 linearly spaced levels ranging from 6 K km s$^{-1}$ to the maximum integrated intensity measured in each 0.5 km s$^{-1}$ velocity interval; the maximum value is indicated in each panel. The scale bar corresponds to 3 pc at a distance of 3.0 kpc.}
\label{fg6}
\end{figure*}

\begin{figure*}
\center
\includegraphics[width=\textwidth]{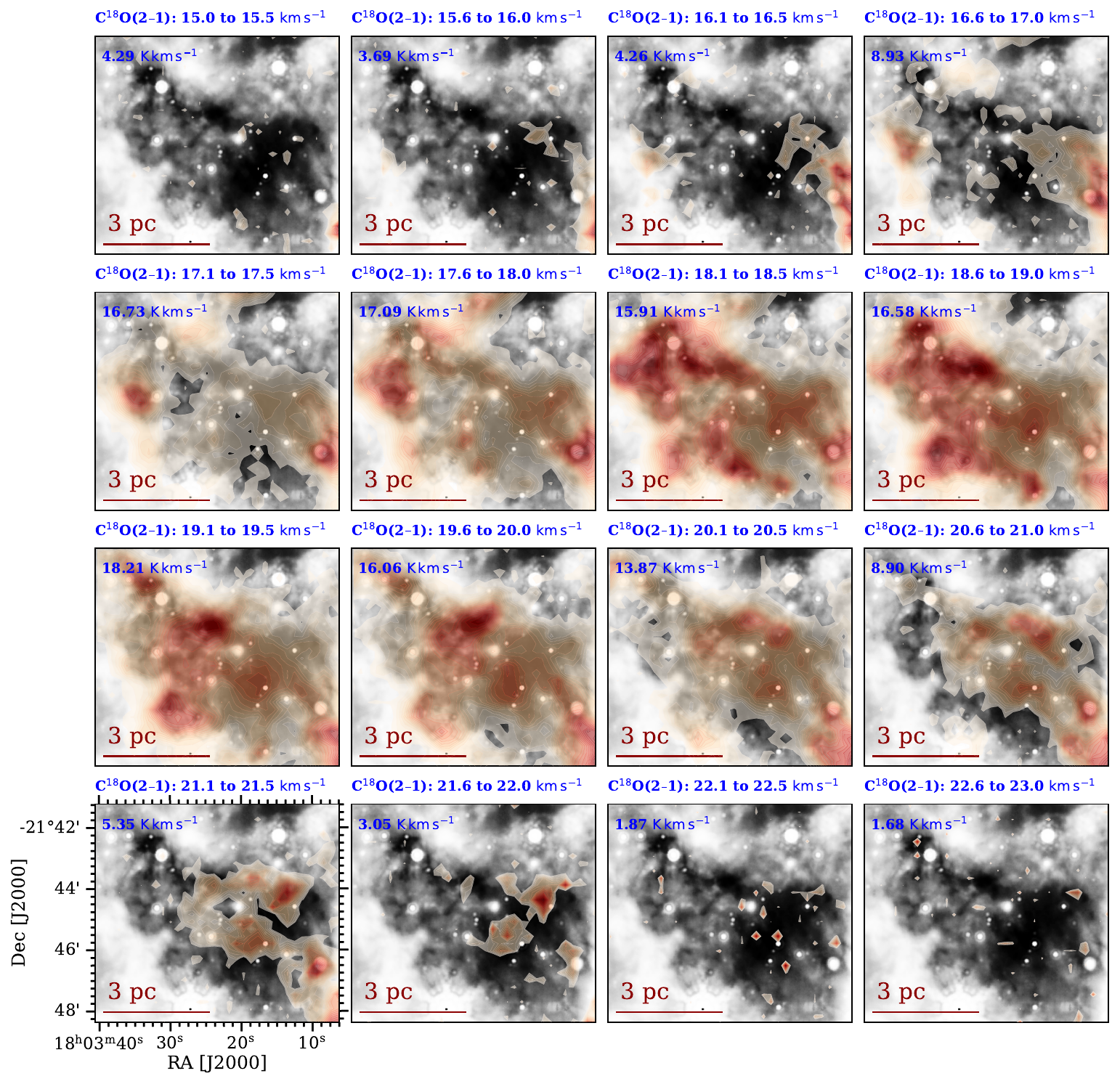}
\caption{Channel maps of the APEX C$^{18}$O(2--1) emission. The C$^{18}$O emission is shown as filled contours, with 20 linearly spaced levels ranging from 6 K km s$^{-1}$ to the maximum integrated intensity in each channel map; the maximum value is indicated in each panel. The channel width is 0.5 km s$^{-1}$. The scale bar corresponds to 3 pc at a distance of 3.0 kpc.} 
\label{fg7}
\end{figure*}

Figures~\ref{fg5}c and~\ref{fg5}d show the spatial distributions of the two velocity components traced by the $^{13}$CO(2--1) and C$^{18}$O(2--1) emission at 18.0 and 20.5~km~s$^{-1}$, respectively. These velocity channels were selected based on the two components identified in the molecular-line spectra (Figure~\ref{fg5}b). Plus symbols shown in Figure~\ref{fg5}a are also marked in Figures~\ref{fg5}c and~\ref{fg5}d. To further explore this aspect, Figures~\ref{fg6} and~\ref{fg7} show the channel maps of the $^{13}$CO(2--1) and C$^{18}$O(2--1) emission, respectively, overlaid on the 24~$\mu$m image. The $^{13}$CO emission exhibits a continuous spatial evolution across the 17.0--21.0~km~s$^{-1}$ velocity range, rather than two completely distinct structures. At 15.0--17.4~km~s$^{-1}$, the emission is relatively localized and fragmented, while between 17.5 and 20.9~km~s$^{-1}$ it becomes more extended and spatially coherent. Toward 21.4--22.9~km~s$^{-1}$, the extended emission progressively weakens and becomes localized again. Thus, the two components identified near 18.1 and 20.5~km~s$^{-1}$ are not spatially disconnected, but exhibit overlapping distributions and a smooth morphological evolution with velocity. A similar spatial and kinematic continuity is also evident in the C$^{18}$O(2--1) emission.

\begin{figure*}
\center
\includegraphics[width=\textwidth]{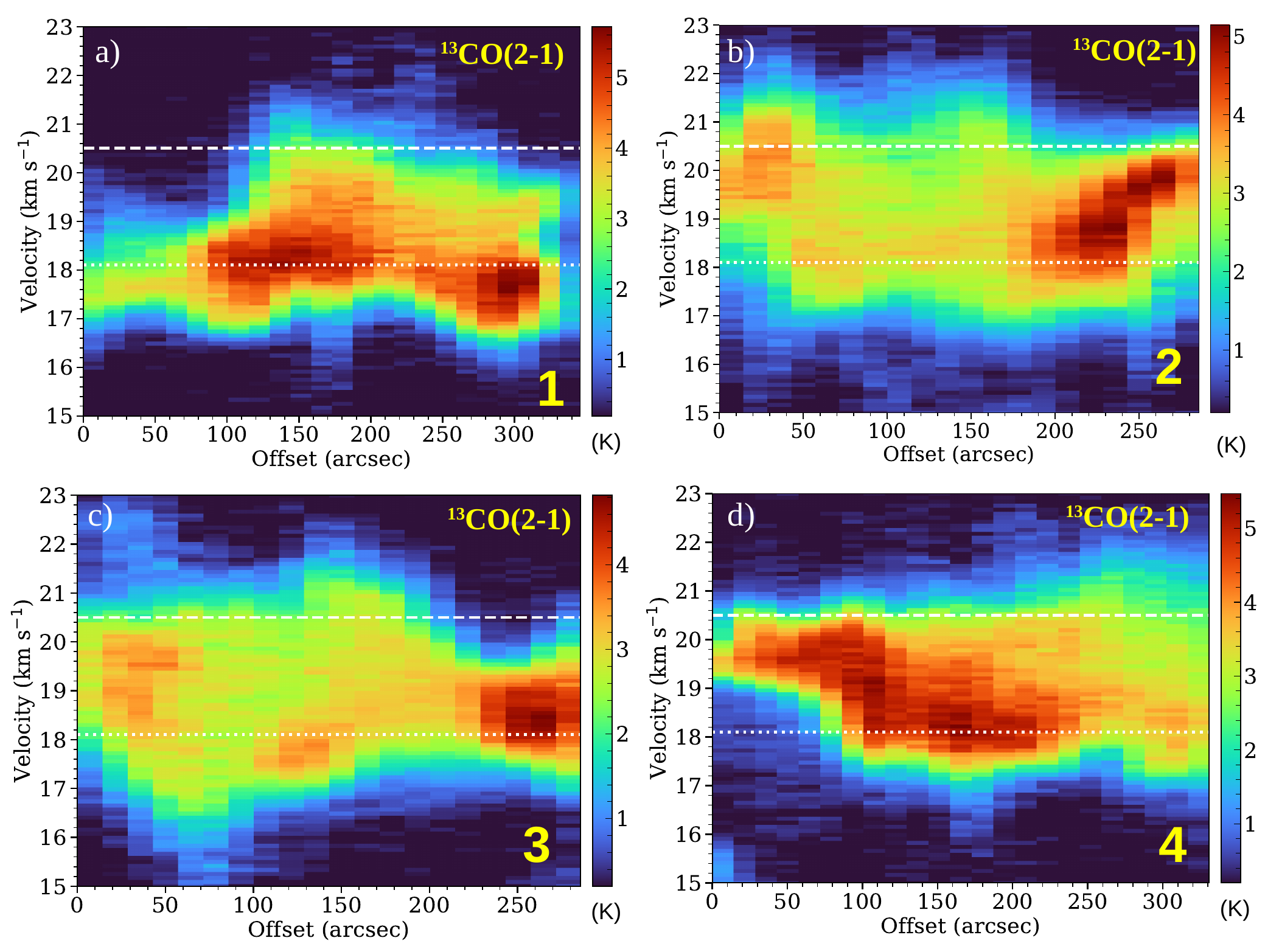}
\caption{a-d) APEX $^{13}$CO(2--1) position-velocity diagrams extracted along the four arrows labeled ``1--4'' in Figure~\ref{fg2}d.} 
\label{fg8}
\end{figure*}

\begin{figure*}
\center
\includegraphics[width=\textwidth]{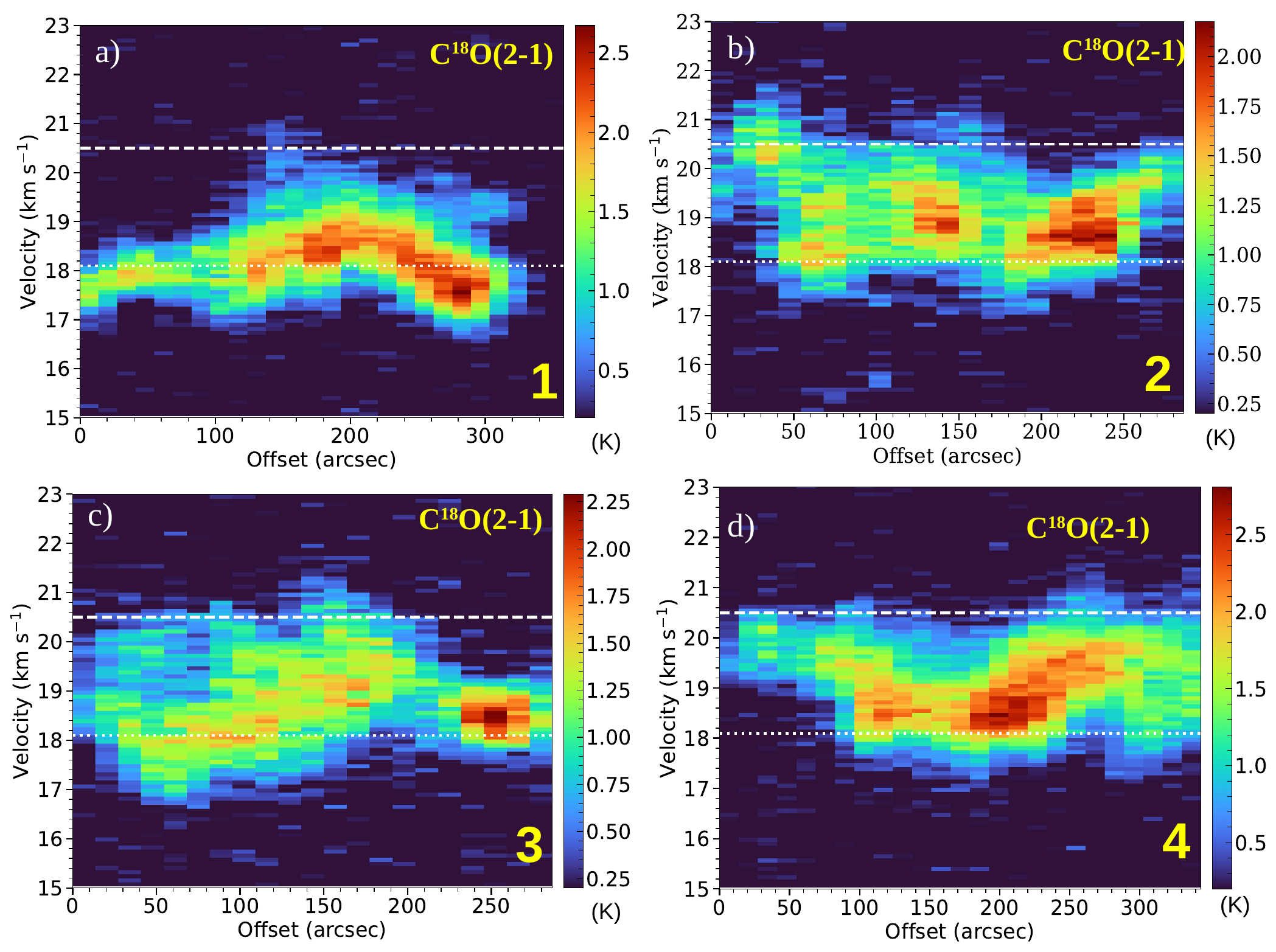}
\caption{Same as Figure~\ref{fg8}, but for the APEX C$^{18}$O(2--1) emission.} 
\label{fg9}
\end{figure*}

The position-velocity diagrams of the $^{13}$CO(2--1) and C$^{18}$O(2--1) emission along four selected paths crossing the dark features within the EFS are presented in Figures~\ref{fg8} and~\ref{fg9}, respectively. The locations of these paths are marked in Figures~\ref{fg2}d and~\ref{fg5}a. Three of the four paths are oriented nearly perpendicular to the EFS spine or curved structure (see Figure~\ref{fg3}b). The position-velocity diagrams show that the two velocity components are not kinematically isolated, but remain spatially connected and partially overlap along the four paths. Their relative contribution varies with position, while their velocity separation remains modest ($\sim$1--2~km~s$^{-1}$). The C$^{18}$O emission traces the denser parts of these structures and shows similar position-velocity continuity. The emission is predominantly distributed over $\sim$18--20~km~s$^{-1}$ and exhibits curved, elongated, and vertically extended structures across the different cuts, rather than a single narrow kinematic component. The systematic shift of emission in velocity with position along several cuts indicates velocity gradients, while the vertically extended emission suggests multiple or overlapping velocity components along the line of sight. Collectively, the position-velocity diagrams reveal non-uniform and complex molecular-gas kinematics, with the two velocity components remaining closely associated across the region.

Taken together, these results reveal at least two velocity components with a small velocity separation, extended position-velocity continuity, and filamentary morphology, suggesting large-scale organization of the molecular gas toward the EFS. The implications of these observed characteristics for molecular-gas assembly and star formation are discussed in Section~\ref{sec:disc}.
\section{Discussion}
\label{sec:disc}
The EFS encompasses two contrasting environments: the BPN associated with the G8.14+0.23 H\,{\sc ii} region and the dark features that lack detectable radio continuum emission. The presence of clusters of YSOs and several clumps with surface densities $>1$~g~cm$^{-2}$ further highlights the EFS as a promising site for investigating MSF and stellar-cluster formation (Figures~\ref{fg1}a--\ref{fg1}c; \citealt{dewangan12,dewangan16b,dewangan19a}). Using the APEX C$^{18}$O-derived H$_2$ column density map, the total molecular mass of the EFS is estimated to be $\sim$10$^5$~$M_{\odot}$ (see Section~\ref{1zzsec:gas}), indicating a substantial molecular reservoir. The BPN's waist hosts a 6.7~GHz MME and the powering source(s) of the G8.14+0.23 H\,{\sc ii} region. In particular, the dark features provide an opportunity to investigate the physical conditions associated with early MSF and possible gas assembly in a comparatively less ionized environment. However, the detailed molecular-gas structure and kinematics of the EFS have remained poorly constrained.

The APEX $^{13}$CO and C$^{18}$O emission toward the dark-feature region exhibits two closely spaced velocity components (see Sections~\ref{sec:gasmorphb},~\ref{sec:moldt}, and Appendix~\ref{sec:posbfield}). To examine their spatial relationship with the EFS, we performed Gaussian decomposition of the $^{13}$CO(2--1) and C$^{18}$O(2--1) line profiles. The histograms of the fitted centroid velocities reveal two distinct peaks, indicating the presence of two dominant velocity components toward the EFS (see Figure~\ref{af1e}). Based on this decomposition, we present maps showing the number of velocity components along each line of sight, the spatial distributions of the individual components over the velocity ranges of 15--19.35 and 19.55--23~km~s$^{-1}$, and their combined spatial distribution toward the EFS (see Figures~\ref{af1b} and~\ref{af1c} in Appendix~\ref{ccsec:posbfield}). The two components exhibit distinct but partially overlapping spatial distributions.
Position-velocity diagrams along four independent paths further reveal coherent velocity structures associated with both components, with increasing spatial association toward their overlapping region in the EFS. Together with the continuous velocity distribution in the channel maps, these characteristics suggest that the components may represent related portions of the larger molecular reservoir, potentially participating in the assembly of the observed molecular structures. Such a configuration could arise from low-velocity converging motions or large-scale gravitational gas transport. Similar velocity-coherent structures have been reported in massive star-forming regions and interpreted as signatures of large-scale gas transport toward dense hubs and condensations \citep[e.g.,][]{peretto13,kirk13,henshaw14,zhou22}. In Mon~R2, the spatial convergence of velocity-coherent structures, increasing velocity overlap toward the central hub, and continuous intermediate-velocity emission have similarly been interpreted as evidence for hierarchical gas transport and converging flows \citep{dewangan26monr2}.

Additionally, the C$^{18}$O(2--1) PPV diagram exhibits an oscillatory velocity pattern over the 19.7--24~km~s$^{-1}$ interval toward the EFS, which hosts several clumps (Figure~\ref{fg5x}b). This pattern is also evident in the PV diagrams in Figures~\ref{fg4}b and~\ref{fg4}c, particularly in the red contours highlighting the undulating velocity structure over offsets of $\sim$1--7~arcmin (Figure~\ref{fg4}c). Such oscillatory velocity patterns have been interpreted as signatures of filament fragmentation and associated gas motions during core formation within filamentary structures \citep{liu19}. However, given the presence of two closely spaced velocity components toward the EFS, the observed pattern may also reflect the superposition of nearby velocity structures along the line of sight, in addition to possible fragmentation-related motions. The relatively small velocity separation of $\sim$2.4~km~s$^{-1}$ is compatible with low-velocity converging motions, although it does not by itself establish convergence or exclude a cloud--cloud collision (CCC) scenario. As discussed by \citet[][see their Figure~11]{Maity_2024}, the velocity separation between parental clouds can decrease substantially at later evolutionary stages, when the original clouds may no longer remain distinguishable and a compressed layer can develop between them. Thus, the present-day velocity separation alone cannot be used to rule out a collision scenario. Taken together, the observed velocity coherence, partial spatial overlap, and oscillatory pattern point to complex molecular-gas motions that may involve fragmentation, gravitational transport, and interactions between spatially associated velocity components. These characteristics are broadly consistent with hierarchical molecular-cloud evolution, in which material is transported from larger-scale structures toward progressively denser clumps, hubs, and cores \citep{Motte+2018,semadeni19}. Nevertheless, the available observations do not uniquely establish the underlying dynamical process or distinguish between gradual gravitational assembly, converging motions, and a possible late-stage CCC. In particular, they do not demonstrate two global flows converging toward a common location or establish the three-dimensional gas dynamics. 

The MIR morphology provides an independent spatial constraint on this interpretation. The unWISE 12~$\mu$m and {\it Spitzer} 24~$\mu$m emission reveals at least three spatially distinct infrared features, with filamentary structures surrounding or extending toward compact infrared emission associated with clusters of YSOs (see Figure~\ref{fg5}a). These features are embedded within the broader molecular structure exhibiting the two partially overlapping velocity components, suggesting that localized star formation occurs within the larger molecular reservoir. Similar configurations of filamentary structures, dense hubs, and compact star-forming sites have been reported in massive star-forming regions \citep[e.g.,][]{peretto13,Motte+2018,zhou22,yang23}. However, the angular resolution of the molecular-line data is insufficient to establish direct kinematic associations between individual velocity structures and the compact infrared features. Consequently, the relationship between the localized star-forming sites and the larger-scale molecular-gas motions remains unresolved.

The observed morphology and kinematics are also qualitatively consistent with recent simulations in which large-scale gas flows organize molecular material into dense filaments that channel gas toward forming hubs \citep{nozaki26}. Although such a scenario provides a possible framework for interpreting the observed molecular structures, the present observations do not resolve individual inflow channels or establish systematic gas convergence toward a common hub.

Overall, the EFS represents a massive molecular environment in which organized molecular structures and localized star formation coexist, while stellar feedback is already influencing the molecular material around the evolved H\,{\sc ii} region. The dark-feature region, in particular, offers an opportunity to investigate the relationship between molecular-gas organization, fragmentation, and star formation in a comparatively less ionized environment. The observed velocity-coherent structures, partial spatial overlap of the velocity components, oscillatory velocity pattern, and MIR morphology collectively suggest that large-scale molecular-gas dynamics may contribute to the formation and evolution of dense clumps and young stellar clusters within the EFS.
\section{Summary and Conclusions}
\label{sec:conc}
We conducted spectral line observations with the APEX toward a previously identified elongated filamentary structure (EFS; extent $>$ 15 pc) that hosts the bipolar nebula (BPN) associated with the site G8.14+0.23/IRAS 17599-2148. This EFS is known to host clusters of YSOs and exhibits ongoing MSF activity. 
The newly acquired APEX $^{12}$CO(2--1), $^{13}$CO(2--1), and C$^{18}$O(2--1) line data (beam size $\sim$25$''$--28$''$ or $\sim$0.36--0.41 pc) are employed to investigate the distribution and kinematics of the molecular gas. 
These observations are complemented by archival multi-wavelength data covering infrared to radio wavelengths, including SARAO MeerKAT 1.3 GHz continuum observations (resolution $\sim$8$''$ or $\sim$0.12 pc) of G8.14+0.23, to investigate the physical processes associated with massive-star and cluster formation in the EFS. Our main results are summarized below:

\begin{enumerate}

\item The new APEX observations trace a filamentary molecular cloud spatially consistent with the EFS. The C$^{18}$O-derived H$_2$ column density yields a total molecular mass of $\sim$10$^5$~$M_{\odot}$, confirming the presence of a substantial molecular reservoir. A total of 22 Hi-GAL clumps are identified within the EFS, including at least five with surface densities exceeding $1~{\rm g~cm^{-2}}$, placing them in the commonly adopted high-surface-density regime associated with MSF. 

\item Two prominent molecular velocity components are identified across the EFS in PPV space. Toward the dark-feature region, they are separated by $\sim$2.4 km s$^{-1}$ and exhibit coherent spatial and kinematic structures, with increasing spatial association toward their overlapping region. These characteristics suggest that the two components may be dynamically related and potentially involved in large-scale molecular-gas assembly. However, the available kinematic information does not uniquely distinguish low-velocity converging motions from the superposition of nearby structures or large-scale gravitational motions.

\item The unWISE 12~$\mu$m and {\it Spitzer} 24~$\mu$m images reveal at least three spatially distinct compact infrared features surrounded or approached by filamentary structures. Their association with YSO clusters and location within the broader molecular structure indicate that localized star formation is embedded within the larger molecular reservoir. However, the APEX angular resolution does not allow individual velocity components to be uniquely associated with these compact star-forming features. 

\item The C$^{18}$O position-velocity diagrams reveal an undulating velocity pattern over offsets of $\sim$1--7 arcmin, suggesting velocity oscillations along the molecular structure, along with a broader velocity spread ($\sim$7--9 km s$^{-1}$) toward the BPN waist. A prominent circular/arc-shaped velocity structure surrounding the H\,{\sc ii} region further supports the presence of an expanding molecular shell, consistent with the previously reported expansion scenario. The molecular environment surrounding the G8.14+0.23 H\,{\sc ii} region thus exhibits kinematic signatures of expansion, indicating that stellar feedback has already influenced the surrounding molecular gas. In contrast, the dark-feature region lacks detectable radio continuum emission and offers an opportunity to investigate molecular-gas structures and star formation in a comparatively less feedback-affected environment.

\item Overall, the EFS exhibits organized molecular-gas structures and localized star formation, while stellar feedback is already influencing the molecular environment around the evolved H\,{\sc ii} region. The observed spatial and kinematic association between the two velocity components is consistent with possible low-velocity gas convergence and large-scale molecular-gas assembly, potentially linking the evolution of the molecular reservoir to the formation of dense clumps and young stellar clusters. However, the present observations do not establish two global converging flows or uniquely constrain the three-dimensional gas dynamics. Higher-angular-resolution molecular-line and dust-continuum observations will be essential to resolve the gas kinematics and investigate their connection to the formation of dense structures and massive stars on smaller spatial scales.
\end{enumerate}
\section*{Acknowledgments}
We sincerely thank the anonymous referee for a careful and detailed review, whose constructive suggestions have led to significant improvements in this manuscript.  
The research work at Physical Research Laboratory is funded by the Department of Space, Government of India. 
ORJ gratefully acknowledges PRL for a Ph.D. research fellowship. IIZ and LEP are supported by the Russian Science Foundation grant 24-12-00153. DKO acknowledges the support of the Department of Atomic Energy, Government of India, under Project Identification No. RTI 4012. AKM acknowledges support from JSPS KAKENHI Grant Number JP26KF0052. This publication is based on data acquired with the Atacama Pathfinder Experiment (APEX) under the programme ID 0104.F-9301(A). 
APEX is a collaboration between the Max-Planck-Institut fur Radioastronomie, the European Southern Observatory, 
and the Onsala Space Observatory. 
This research has made use of the NASA/IPAC Infrared Science Archive, which is funded by the National Aeronautics and Space Administration and operated by the California Institute of Technology. This research made use of {\it Astropy}\footnote[4]{http://www.astropy.org}, a community-developed core Python package for Astronomy \citep{astropy13,astropy18}. For figures, we have used {\it matplotlib} \citep{Hunter_2007} and IDL software.  
\subsection*{Data availability}
The {\it Spitzer}-MIPSGAL data underlying this article are available from the publicly accessible NASA/IPAC infrared science archive\footnote[5]{https://irsa.ipac.caltech.edu/data/SPITZER/MIPSGAL/overview.html}. 
The unWISE data underlying this article are available from the publicly accessible website\footnote[6]{https://unwise.me}.
The Hi-GAL clumps underlying this article are available from the publicly accessible website\footnote[7]{https://cdsarc.cds.unistra.fr/viz-bin/cat/J/MNRAS/471/100}.
The APEX $^{12}$CO(2--1), $^{13}$CO(2--1), and C$^{18}$O(2--1) line data underlying this article are available from the publicly accessible APEX instrument specific query form\footnote[8]{https://archive.eso.org/wdb/wdb/eso/apex/form/}.
The ATLASGAL 870 $\mu$m dust continuum clumps underlying this article are available from the publicly accessible website\footnote[9]{https://cdsarc.cds.unistra.fr/viz-bin/cat/J/MNRAS/473/1059}.
The SARAO MeerKAT 1.3 GHz continuum data underlying this article are available from the publicly accessible website\footnote[10]{https://archive-gw-1.kat.ac.za/public/repository/10.48479/3wfd-e270/index.html}.
%
%


\bibliographystyle{mnras}
\bibliography{reference} 
\appendix
\section{Channel maps of APEX $^{13}$CO(2--1)}
\label{sec:posbfield} 
\begin{figure*}
\center
\includegraphics[width=\textwidth]{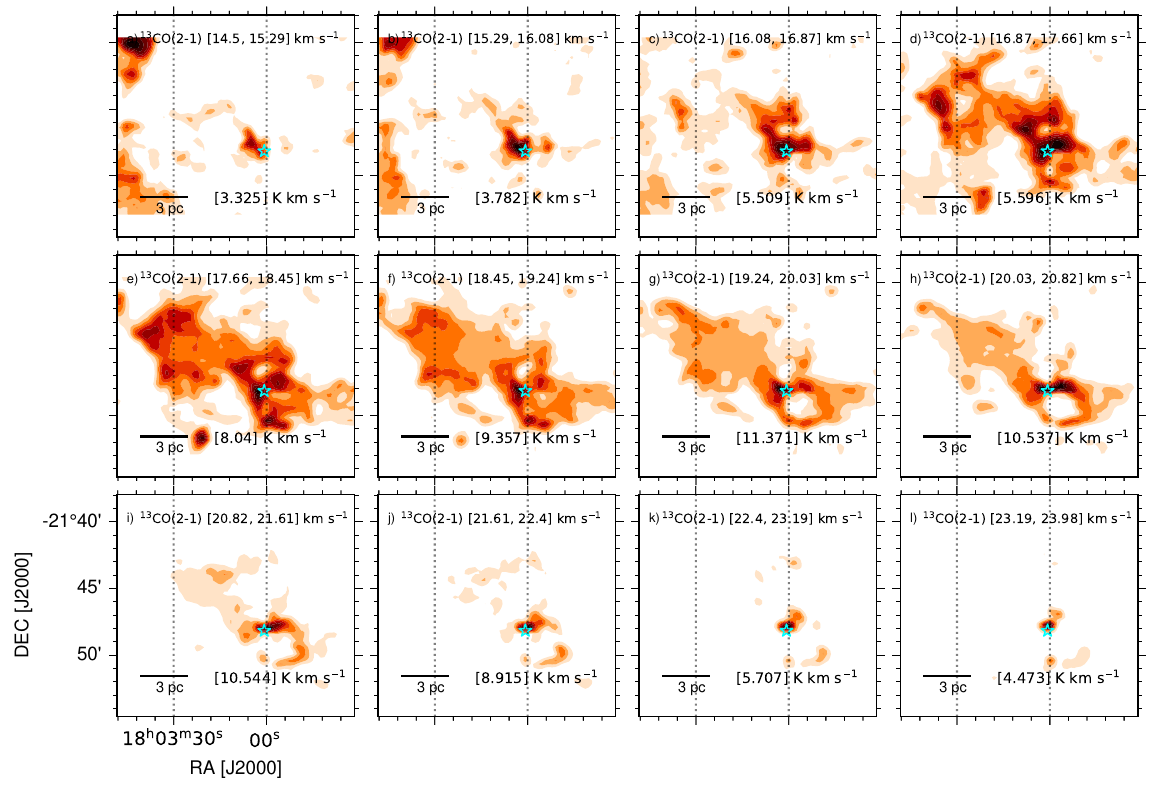}
\caption{Channel maps of the APEX $^{13}$CO(2--1) emission for the area shown in Figure~\ref{fg2}a. Filled contours in each velocity channel correspond to [0.1, 0.2, 0.3, 0.4, 0.5, 0.6, 0.7, 0.8, 0.9, 1.0] $\times$ the peak intensity of that channel, with the peak value given in each panel. The star symbol shows the position of the 6.7 GHz MME \citep[from][]{szymczak12}, 
and the scale bar corresponds to 3 pc at a distance of 3.0 kpc.} 
\label{af1}
\end{figure*}

In Figure~\ref{af1}, we present the channel maps of the APEX
$^{13}$CO(2--1) emission. The molecular gas exhibits pronounced
velocity-dependent morphological variations toward EFS. At lower
velocities ($\sim$14.5--18.5~km~s$^{-1}$), an extended molecular
component is prominent toward the eastern side of the EFS, coinciding
with the infrared-dark regions where no radio continuum emission is
detected. Notably, in the velocity interval $\sim$16.9--18.5~km~s$^{-1}$,
the molecular emission toward the eastern side forms an almost circular,
shell-like morphology, suggestive of molecular gas surrounding a
lower-column-density cavity (see panels ``d'' and ``e'' in Figure~\ref{af1}). 
With increasing velocity, the molecular emission
becomes progressively concentrated toward the waist of the BPN, where
the 6.7~GHz MME and radio continuum peaks are located, and extends
toward the southwestern side. The continuous spatial evolution of the molecular emission across the velocity channels suggests the presence of two kinematically connected molecular structures.
\section{Spectral Decomposition}
\label{ccsec:posbfield} 
We performed Gaussian decomposition of the $^{13}$CO(2--1) and C$^{18}$O(2--1) molecular line data using the \textit{scousepy} package. The decomposition was performed using the four-stage fitting procedure implemented in scousepy, whereby multiple Gaussian components were fitted to the individual spectra at each spatial pixel across the observed field. For each fitted component, we extracted the peak intensity, centroid velocity, and velocity dispersion ($\sigma_v$).

\begin{figure*}
\center
\includegraphics[width=\textwidth]{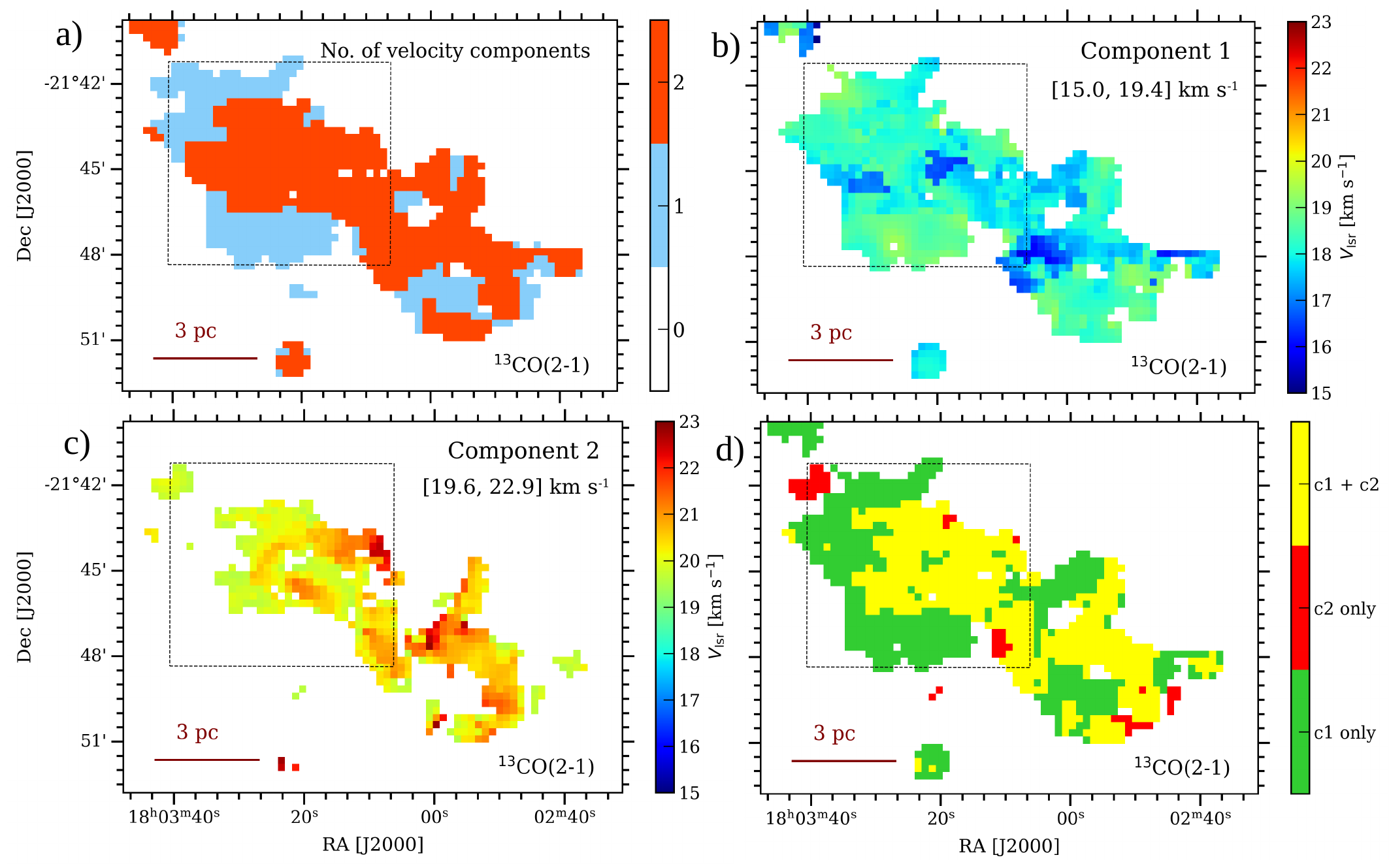}
\caption{a) Number of velocity components identified in the $^{13}$CO(2--1) emission. b) Centroid velocity map of component 1 (c1; $V_{\rm lsr}<19.35$ km s$^{-1}$). c) Centroid velocity map of component 2 (c2; $V_{\rm lsr}>19.55$ km s$^{-1}$). d) Spatial distribution of the two velocity components, highlighting their regions of overlap. Green indicates pixels containing only component 1, red indicates pixels containing only component 2, and yellow indicates pixels where both components are detected. In all panels, the dashed box highlights the area containing the infrared dark features (see also Figure~\ref{fg1}a). In each panel, the scale bar corresponds to 3~pc at a distance of 3.0 kpc.}
\label{af1b}
\end{figure*}

\begin{figure*}
\center
\includegraphics[width=\textwidth]{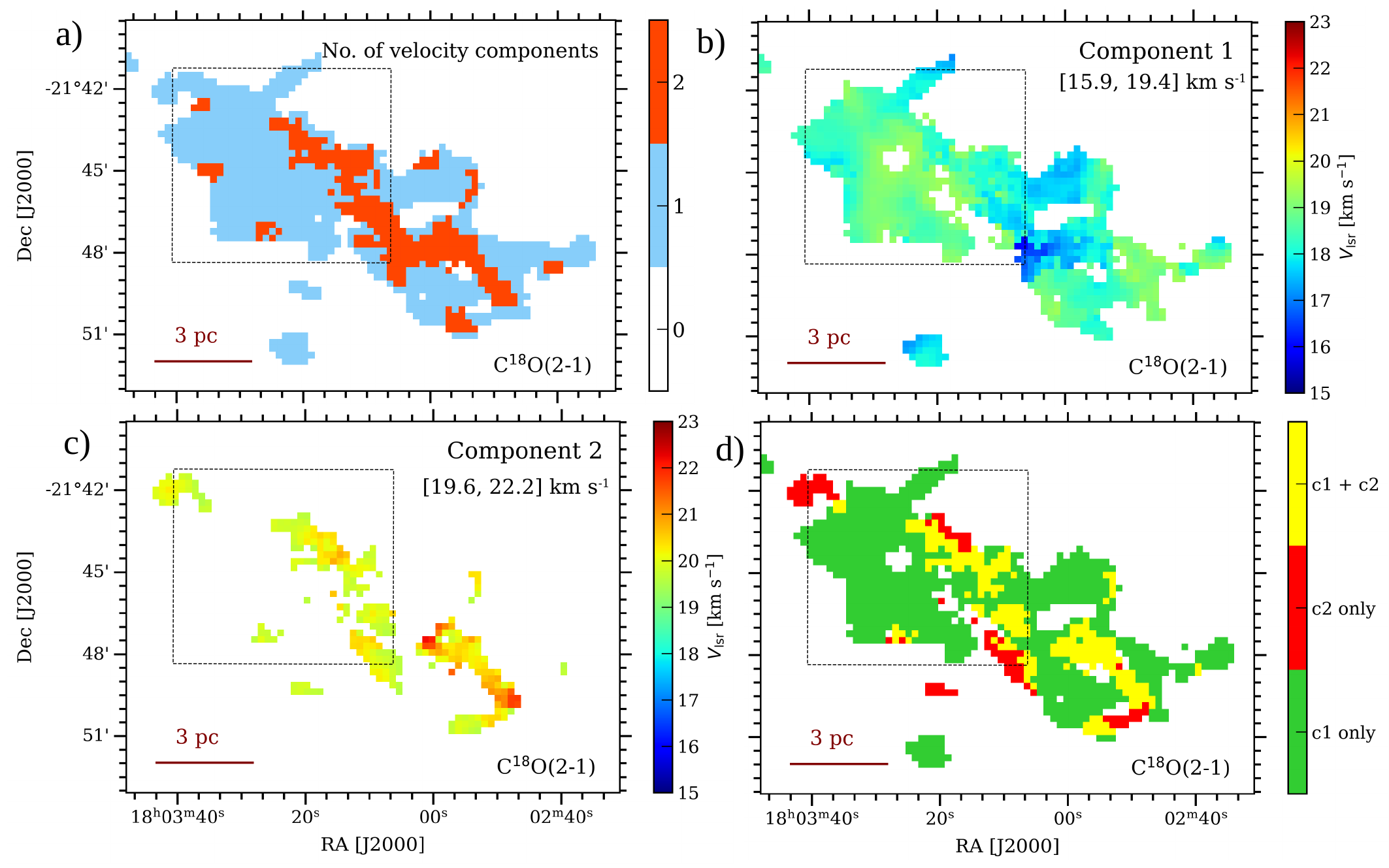}
\caption{a) Number of velocity components identified in the C$^{18}$O(2--1) emission. b) Centroid velocity map of component 1 (c1; $V_{\rm lsr}<19.35$ km s$^{-1}$). c) Centroid velocity map of component 2 (c2; $V_{\rm lsr}>19.55$ km s$^{-1}$). d) Spatial distribution of the two velocity components, highlighting their regions of overlap. Green indicates pixels containing only component 1, red indicates pixels containing only component 2, and yellow indicates pixels where both components are detected. In all panels, the dashed box highlights the area containing the infrared dark features (see also Figure~\ref{fg1}a). In each panel, the scale bar corresponds to 3~pc at a distance of 3.0 kpc.}
\label{af1c}
\end{figure*}

The analysis was restricted to the velocity range of 15-23 km s$^{-1}$. Based on the fitted centroid velocities, we selected components with $V_{\rm lsr}<19.35$ km s$^{-1}$ to represent component 1 and with $V_{\rm lsr}>19.55$ km s$^{-1}$ to represent Component 2 (see also Figure~\ref{fg5}b). These two velocity limits define the boundaries of the gray-shaded interval encompassing the intersection velocity ($\sim$19.45 km s$^{-1}$) of the two fitted Gaussian profiles in Figure~\ref{fg5}b. We subsequently examined the spatial distribution of the fitted components for $^{13}$CO(2--1) and C$^{18}$O(2--1) data and generated a pixel-by-pixel map showing the number of velocity components identified across the region (Figures~\ref{af1b}a and~\ref{af1c}a). Figures~\ref{af1b}b and~\ref{af1b}c show the velocity structure of the two $^{13}$CO(2--1) velocity components (Components 1 and 2), while Figures~\ref{af1c}b and~\ref{af1c}c show the corresponding velocity structure traced by C$^{18}$O(2--1). The spatial overlap between the two velocity components (i.e., [15, 19.35] and [19.55, 23]~km~s$^{-1}$) is shown in Figures~\ref{af1b}d and~\ref{af1c}d for $^{13}$CO(2--1) and C$^{18}$O(2--1), respectively.

\begin{figure*}
\center
\includegraphics[width=\textwidth]{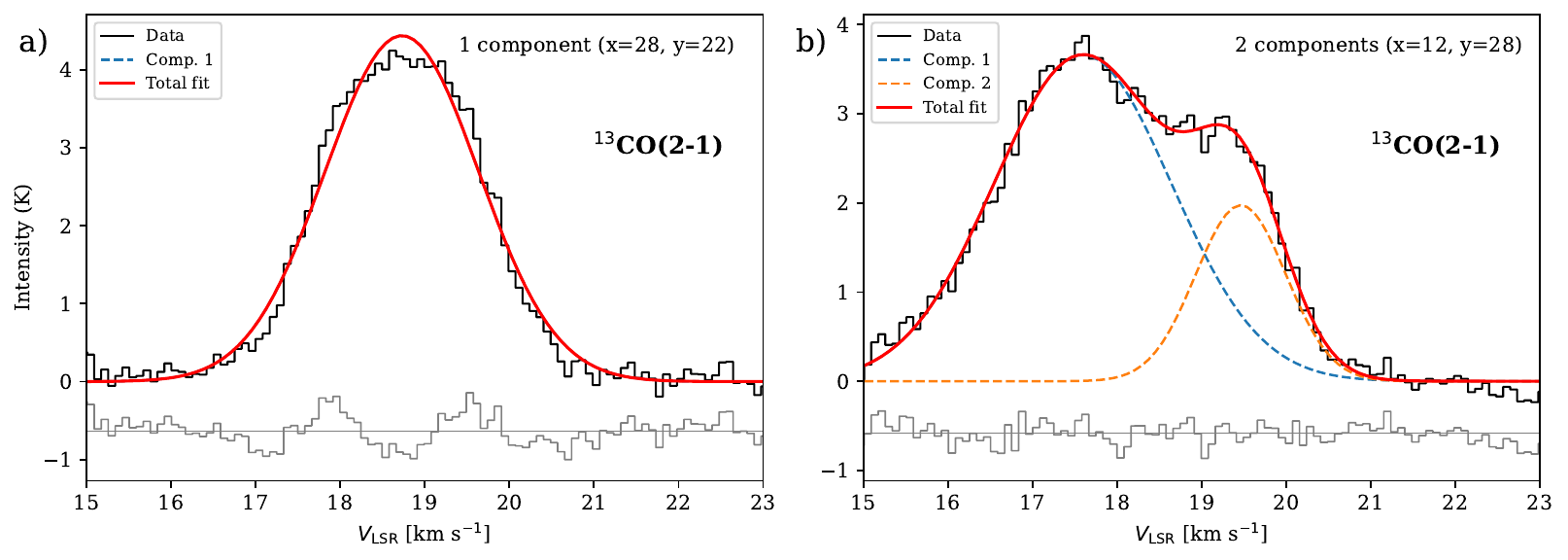}
\caption{a) Observed $^{13}$CO(2--1) spectrum (black) at pixel (28, 22), fitted with a single Gaussian component (red). b) Observed $^{13}$CO(2--1) spectrum (black) at pixel (12, 22), fitted with two Gaussian components (blue and orange); their combined fit is shown in red. The residuals of the fits are shown in grey in both panels.}
\label{af1d}
\end{figure*}

\begin{figure*}
\center
\includegraphics[width=\textwidth]{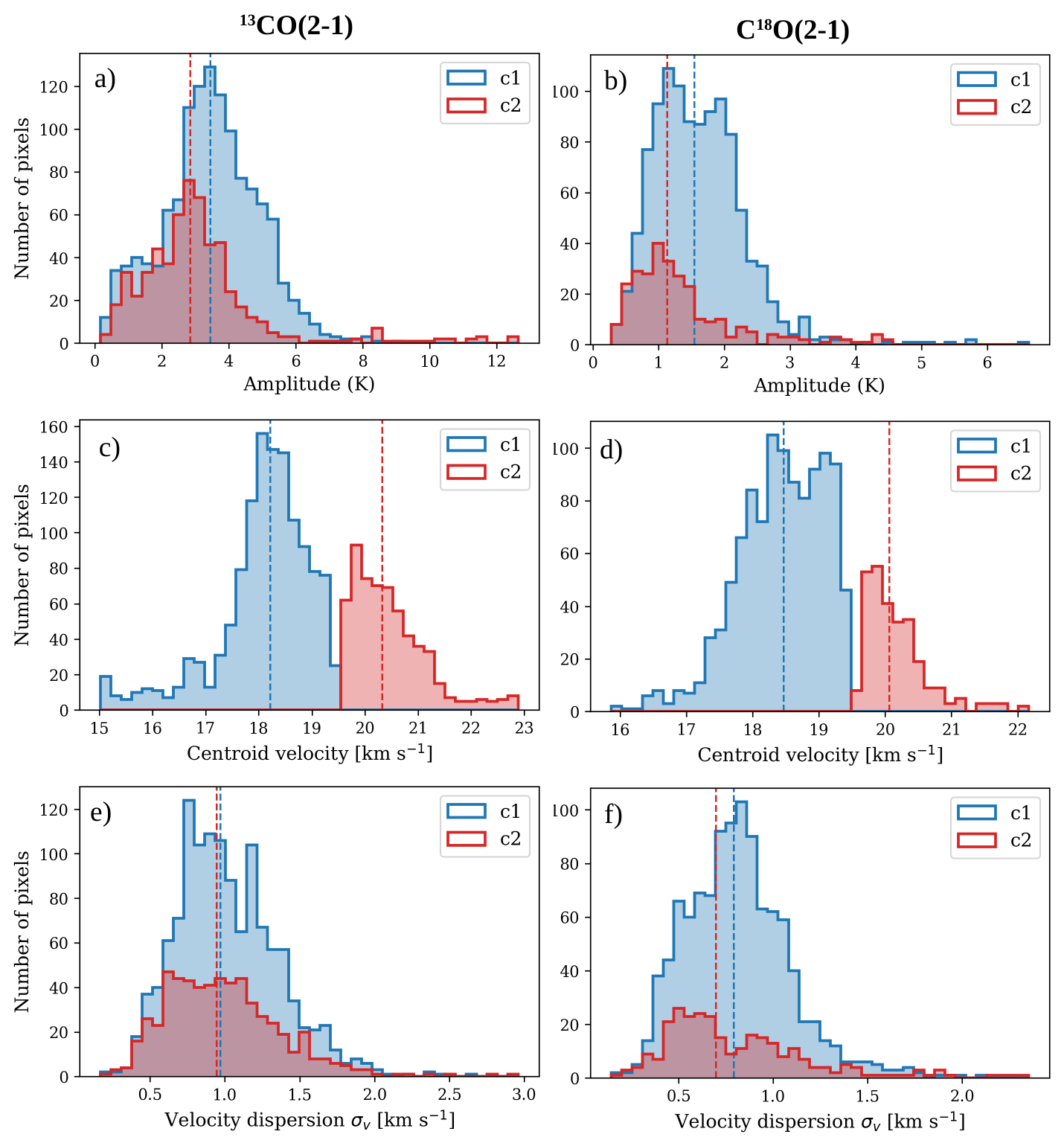}
\caption{Panels (a), (c), and (e) show the histograms of the fitted amplitudes, centroid velocities, and velocity dispersions, respectively, for the two $^{13}$CO(2--1) velocity components. Panels (b), (d), and (f) show the corresponding histograms for the two C$^{18}$O(2--1) velocity components.}
\label{af1e}
\end{figure*}

Figure~\ref{af1d} shows representative spectra from two pixels, illustrating examples of the single- and double-Gaussian fitting applied to the $^{13}$CO(2--1) data. Figure~\ref{af1e} shows the histogram distributions of the fitted parameters for the two velocity components (components 1 and 2). Both components peak at relatively low amplitudes ($\sim$1 K), although component 1 is brighter, with a median amplitude of $\sim$1.5 K compared to $\sim$1.1 K for component 2, and exhibits a tail extending beyond 6 K. The two components are clearly separated in velocity, with median velocities of $\sim$18.5 km s$^{-1}$ (component 1) and $\sim$20.0 km s$^{-1}$ (component 2). The sharp edges near 19.5 km s$^{-1}$ arise from the adopted velocity cut. Both components have comparable velocity dispersions. 
The double-peaked spectra are more appropriately described by two distinct velocity components rather than a single broadened line.
\end{document}